\documentclass[prb,aps]{revtex4}
\usepackage{graphicx}
\usepackage{amsmath}

\begin{document}
\title{Magnetostriction in an array of spin chains under magnetic field}
\author{E. Orignac}
\affiliation{Laboratoire de Physique Th\'eorique de l'\'Ecole Normale
  Sup\'erieure CNRS UMR8549 24, Rue Lhomond 75231 Paris Cedex 05
  France}
\author{R. Citro}
\affiliation{ Dipartimento di Fisica ``E. R. Caianiello'' and
Unit{\`a} I.N.F.M. di Salerno\\ Universit{\`a} degli Studi di Salerno, Via S.
Allende, I-84081 Baronissi (Sa), Italy}
\date{\today}
\begin{abstract}
We consider an array of XX spin-1/2 chains coupled to acoustic phonons
and placed in a magnetic field. Treating the phonons in the mean
field approximation, we show that this system presents a first
order transition as a function of the magnetic field 
between a partially magnetized distorted state and the
fully polarized undistorted state at low temperature. 
This behavior results
from the magnetostriction of the coupled chain system. A dip in
the elastic constant of the material near the saturation field
 along with an anomaly in the magnetic
susceptibility is predicted. We also predict the
contraction of the material as the magnetic field is reduced
(positive magnetostriction) and the reciprocal effect i.e. a decrease
of magnetization under applied pressure.  
At higher temperature, the first order transition is replaced by a
crossover. However, the anomalies in the susceptibilities in the
system near the saturation field are still present. 
We discuss the relevance of our analysis in relation to recent
experiments on spin-1/2 chain and ladder materials in high magnetic fields.
\end{abstract}
\maketitle

\section{Introduction}
\label{sec:intro} 
The subject of Bose-Einstein Condensation (BEC) has become
tremendously active both experimentally and theoretically in the
recent years due to the observation of BEC in trapped atomic
gases.\cite{dalfovo99_bec_review} Another way of obtaining a BEC, predicted theoretically some time
ago, is to place a  quasi one dimensional spin gap systems under
a strong magnetic
field\cite{affleck_field,tsvelik_field,chitra_spinchains_field,furusaki_dynamical_ladder,usami_numerics_ladder,konik_spinfield,lou00_magnetization_spin1,hikihara_xxz} 
causing the formation of a Luttinger liquid of magnons. This Luttinger
liquid possesses a quasi-long range order, which is a one dimensional
precursor of the BEC.\cite{haldane_bosons,cazalilla_1d_bec,petrov04_bec_review} In the presence of any
three dimensional coupling, this quasi-condensed state develops a
long-range order, which corresponds to a BEC of the
magnons.\cite{giamarchi_coupled_ladders,ho04_deconfined_bec} 
This condensation has been observed experimentally
\cite{katsumata_nenp_field} in spin-1 chain materials, and more
recently\cite{chaboussant_cuhpcl,chaboussant_nmr_diaza,chaboussant_mapping}
in Cu$_2$(C$_5$H$_{12}$N$_2$)$_2$Cl$_4$ (CuHpCl), 
a material originally believed on the basis of its thermodynamics 
to be formed of weakly
coupled two leg ladders. However, it is now known that CuHpCl is
in fact formed of dimers coupled in a three-dimensional
network\cite{stone02_cuhpcl}, and that the Bose
condensation is a condensation of a two or three-dimensional magnon Bose
fluid.\cite{rice02_bec_magnons} Recently, such a Bose-Einstein condensation of magnons has also been observed in other coupled dimers
systems such as $\mathrm{TlCuCl_3}$\cite{nikuni00_tlcucl3_bec},
$\mathrm{KCuCl_3}$\cite{oosawa02_kcucl3},
$\mathrm{Cs_3Cr_2Br_9}$\cite{grenier04_cr3cr2br9} and
$\mathrm{BaCu_2Si_2O_6}$.\cite{jaime04_bacu2si2o6} 
Theoretical description taking into account the
two-dimensional nature of these gapped systems have been
developed\cite{wessel01_spinliquid_bec,matsumoto02_tlcucl3,matsumoto04_tlcucl3,misguich04_tlcucl3}
in recent years. They are based on the bond operator theory
(BOT) description of the coupled dimer
system.\cite{chubukov89,sachdev_bot,gopalan_2ch}  The interactions
between the condensed triplet bosons are treated by the Hartree-Fock-Popov (HFP)
approximation\cite{popov_functional_book}, and a rigid lattice  is
assumed. However, more recently, some experimental evidence for a
lattice distortion associated with the Bose Einstein Condensation has
been obtained.\cite{sherman03_tlcucl3,vyaselev04_tlcucl3,lorenzo04_cuhpcl_spinpeierls}
Older reports of specific heat anomalies also pointed out to a
lattice distortion associated with the
transition.\cite{calemczuk_heat_ladder}
 It has thus been suggested that the  distortion observed in experiments
could be the generalized spin-Peierls
transition predicted originally for spin
ladder systems under field.\cite{nagaosa_lattice_ladder} 
However, similarly to the
conventional spin-Peierls
transition\cite{pincus_spin-peierls,cross_spinpeierls}, in a
generalized spin-Peierls transition  the  lattice
distortion results from an \emph{optical} phonon becoming static. 
As a result, in the generalized spin-Peierls transition,  the lattice
constants of  the crystal do not change, but superlattice peaks are
visible in X-Ray diffraction, in contrast with
experiments.\cite{lorenzo04_cuhpcl_spinpeierls,vyaselev04_tlcucl3}
In fact, in the case of a condensate of magnetic
excitations coupled to a deformable lattice, another instability than
the generalized spin-Peierls transition is possible, namely
magnetostriction. Although
magnetostriction has been mostly discussed in the context of
ferromagnetism\cite{bozorth51_ferromagnetism,vonsovskii74_magnetism2}
it can also be observed in paramagnetic or antiferromagnetic
systems.\cite{bates39_magnetism} 
In the case of magnetostriction, the lattice
parameters change as the system becomes magnetized, 
but the number of atoms per unit cell
does not change, which means that no new reflection appear in
X-Ray diffraction experiments.\cite{lorenzo04_cuhpcl_spinpeierls} 
In a ferromagnet,  magnetostriction is
known to turn the second order Curie transition into a first order
transition.\cite{grazhdankina69_1storder} Thus, the magnetostrictive
effects may also explain the first order transition as a function of
magnetic field observed in magnon Bose-Einstein condensation
experiments.\cite{lorenzo04_cuhpcl_spinpeierls,sherman03_tlcucl3,vyaselev04_tlcucl3}
Besides the Bose-Einstein condensation of magnons in spin gap systems,  
magnetostriction effects are also relevant near the saturation field
in the 
spin-1/2 chain material
Cu(II)-2,5bis(pyrazol-1-yl)-1,4-dihydroxybenzene (CuCCP).\cite{wolf_sound_anomaly_spinchain} In fact,
the two effects can be closely related as a ladder system under strong
magnetic field can be mapped for strong rung coupling on an anisotropic
spin-1/2 chain.\cite{mila_ladder_strongcoupling} 
In the present paper, we
will discuss a model of magnetostriction in  1D XX spin chains
coupled by 3D acoustic phonons.  We will study the effect of phonon
coupling at the mean field level and show that the Bose Einstein
Condensation transition of the magnons becomes a first order
transition at low temperature. The origin of the first order character
of the transition can be traced to the hardcore repulsion of the
magnons, which prevents the collapse of
the condensate. Treatments based on
Hartree-Fock-Popov mean field theory neglect the hardcore constraint
and thus fail in the regime where  the first order transition is
obtained. This stresses that a
correct treatment of the hardcore constraint is necessary to describe 
the first order transition observed in
experiments.\cite{sherman03_tlcucl3}   

The plan of the paper is the following. In Sec.\ref{sec:model} we
introduce a model of an array of two leg ladders coupled with
classical phonons as a minimal model for describe the BEC in a
system of hard core bosons coupled to a lattice deformation. From
this model we deduce an effective XXZ spin chain Hamiltonian that
can be rewritten in terms of a free fermion Hamiltonian by the
Jordan-Wigner transformation. In Sec.\ref{sec:zeroT} we focus on
the zero temperature case. By a mean field theory,  we discuss the
first order phase transition at varying the external magnetic
field and the pressure. In particular, we derive a Ginzburg-Landau
expansion of the free energy valid for a very stiff lattice which
allows an analytical discussion of the behavior of the
magnetization and of the lattice parameter as a function of the
magnetic field. We also calculate and discuss the behavior of
various susceptibilities in the distorted case, such as the
elastic modulus, the magnetic susceptibility and the magnetostriction
coefficient  that all present anomalies at the transition.
In Sec.\ref{sec:finiteT} we turn to the positive temperature case
and deduce first a Ginzburg-Landau expansion which allows us to
predict the temperature above which the first order transition is
replaced by a crossover. All the
susceptibilities are then recalculated at finite temperature and some
results compared to recent
experiments.\cite{wolf_sound_anomaly_spinchain} 
We also consider the behavior
of the dilatation coefficient and of the specific heat. Finally in Sec.
\ref{sec:conclusions} we give the conclusions and perspectives for
future work.

\section{Model}
\label{sec:model}

In order to grasp the origin of the first order character of the
BEC in a spin gap system under field coupled to acoustic phonons, 
we want to build a model that can be solved
with a minimum of approximation. We will start with a  an
array of two leg ladders coupled with classical phonons. Its
Hamiltonian reads:
\begin{eqnarray}
  \label{eq:ladder-array}
  H= \sum_{n,m} (1+\eta (u_{n,m}-u_{n+1,m}))[J_\parallel (S_{n,m}^x
   S_{n+1,m}^x +S_{n,m}^y S_{n+1,m}^y) + J_\parallel^z S_{n,m}^z
   S_{n+1,m}^z] 
   + J_\perp \sum_{n,m}
  \mathbf{S}_{n,2m}\cdot\mathbf{S}_{n,2m+1} +
  \frac{\bar{K}}{2}(u_{n,m}-u_{n+1,m})^2 -H \sum_{n,m} S_{n,m}^z ,
\end{eqnarray}
\noindent where $J_{\parallel},J_\parallel^z,J_{\perp}>0$ are the exchange
constants, $K$ is the elastic constant, and $\eta$ is a spin-phonon
coupling constant. We have neglected the kinetic energy of the
phonons, as is done usually in treatments of
magnetostriction.\cite{mattis63_magnetostriction} In the case of a
strong rung interaction $J_\perp$, it is possible to reduce the size
of the Hilbert space by considering only the singlet rung state and
the triplet state of spin $S^z=+1$ and derive an effective XXZ spin
chain
theory.\cite{chaboussant_mapping,mila_ladder_strongcoupling,giamarchi_coupled_ladders}  
The resulting
effective Hamiltonian reads:
\begin{eqnarray}
  \label{eq:ladder-strongcoupl}
  H=J_\parallel \sum_{n,m} (1+\delta_{n,m}) (T_{n,m}^x T_{n+1,m}^x
  +T_{n,m}^y T_{n+1,m}^y+ \Delta T_{n,m}^z T_{n+1,m}^z) +
  \frac{K}{2} \delta_{n,m}^2 - h' T_{n,m}^z,
\end{eqnarray}
\noindent where the pseudospin operator is acting on the reduced
Hilbert space, $\delta_{n,m}=\eta (u_{n+1,2m}-u_{n,2m})$ (we are
assuming $u_{n+1,2m}-u_{n,2m}=u_{n+1,2m+1}-u_{n,2m+1}$ as we are
only considering magnetostriction effects), $K=\bar{K}/\eta^2$,
$\Delta=J_\parallel^z/(2J_\parallel)$, 
and $h'=H-J_\perp-J_\parallel^z/2$ is an
effective magnetic field.\cite{mila_ladder_strongcoupling} The
magnetization per rung of the ladder system is $\tilde{m}=\langle
(S_{n,2m}^z+S_{n,2m+1}^z)\rangle =\langle T_{n,m}^z\rangle +1/2$.  
 In the case of  $SU(2)$ invariant spin ladders
 ($J_\parallel=J_\parallel^z$), one
recovers $\Delta=1/2$.\cite{mila_ladder_strongcoupling} The case of
CuCCP is  also described by the model defined by
Eq.~(\ref{eq:ladder-strongcoupl}) with $\Delta=1$ and the
magnetization given by $\langle T_n^z\rangle$.    
It is interesting to note that the operators
$T_{n,m}^{\pm}=(T_{n,m}^x\pm iT_{n,m}^y)$ satisfy the same commutation
relation as hardcore boson creation and annihilation
operators.\cite{schultz_1dbose,fisher_boson_loc} As a result, The
Hamiltonian~(\ref{eq:ladder-strongcoupl}) also describes a system of
hard-core bosons with nearest neighbor repulsion and a boson-phonon
interaction. The magnetic field $h'$ plays the role of the chemical
potential of the hardcore bosons. When $h'$ is sufficiently negative, 
the density of bosons is zero. In the original problem described by
Eq.~(\ref{eq:ladder-array}) this
corresponds to the spin gap state obtained for $H\ll J_\perp$. When
$h'$ is less negative, the density of hardcore bosons in the ground
state becomes nonzero. As our system is one-dimensional, this leads to
the formation of a quasi-Bose condensate of hard-core bosons, in which
the order parameter of the Bose condensation $T^{\pm}$ possesses
quasi-long range order in the ground state. Hence, the transition
between the spin gap state and the magnetized state in coupled dimer
systems\cite{nikuni00_tlcucl3_bec} can be viewed as the Bose
condensation of hard core triplet bosons. In the case we are
considering, we have to consider the effect of the phonons on this
Bose condensation. Let us first show that the HFP approximation can be
inadequate to deal with the magnetostriction effects resulting from
the spin-phonon interaction\cite{matsumoto04_magnetoelastic}. 
Indeed, if we consider bosons with an on-site repulsion
described by the Hamiltonian:
\begin{eqnarray}
  \label{eq:bose-Hubbard}
  H=\sum_{\langle i,j\rangle} \left[-t(1+\delta_{ij}) b^\dagger_i b_j +\frac
  K 2 \delta_{ij}^2 \right] -\mu \sum_i
  b^\dagger_i b_i + U \sum_i b^\dagger_i b_i( b^\dagger_i b_i-1) + V
  \sum_{\langle i,j\rangle} b^\dagger_i b_i b^\dagger_j b_j ,  
\end{eqnarray}
we notice that in one dimension and in the limit $U\to\infty$, the
Hamiltonian (\ref{eq:bose-Hubbard}) becomes equivalent to the
Hamiltonian~(\ref{eq:ladder-strongcoupl}) 
with $J=t/2$, $V=J\delta$, and $\mu=h'$. 
Let us make  make the HFP approximation $b_i \to \langle
b_i\rangle=\lambda$, $\delta_{ij}\to \delta $ in
(\ref{eq:bose-Hubbard}). 
The resulting ground state energy reads:
\begin{eqnarray}
  \label{HFP-approx}
E&=&N\left[-zt(1+\delta)|\lambda|^2 -(\mu+U) |\lambda|^2 + (U+zV)|\lambda|^4
  +\frac K 2 \delta^2\right],\nonumber \\
 &=&N\left[ -(\mu+U+zt) |\lambda|^2 + \left(U+zV-\frac{z^2 t^2}{2K}\right)
  \left(|\lambda|^2 -\frac{\mu+U+zt}{2\left(U+zV-\frac{z^2
  t^2}{2K}\right)}\right)^2  + \frac K 2 \left(\delta -\frac
  {zt}{K}|\lambda|^2\right)^2\right]+\text{Ct},   
\end{eqnarray} 
where $z$ is the coordination number of the lattice. 
For $U+zV-\frac{z^2 t^2}{2K}>0$, the variational energy
Eq.~(\ref{HFP-approx})  is bounded from below,   
and the minimization of (\ref{HFP-approx}) leads to:
\begin{eqnarray}
  |\lambda|^2&=&\frac{\mu+U+zt}{2U +2zV -\frac {t^2}{K}}\Theta(\mu+U+zt), \label{HFP-mag} \\ 
  \delta &=& \frac {zt} K \frac{\mu+U+zt}{2U + 2zV -\frac {t^2}{K}}\Theta(\mu+U+zt), \label{HFP-distort}
\end{eqnarray}
\noindent where $\Theta(x)$ is the Heaviside step function. 
 From Eqs.~(\ref{HFP-mag})-(\ref{HFP-distort}), we obtain a second
 order phase transition as $\mu>\mu_c^{(HFP)}=-zt-U$. For
 $U+zV-\frac{z^2 t^2}{2K}<0$, the variational energy
 (\ref{HFP-approx}) becomes unbounded from below. In that case, the
 HFP approximation leads to un unphysical result $|\lambda|=\infty$.     
The origin of this unphysical behavior when the spin-phonon
 interaction becomes too strong is easily seen. The HFP approximation 
 reduces the
spin-phonon coupling to an effective attractive interaction between
 the bosons of order
$t^2/K$, and when this attraction overcomes the repulsion $U+zV$, the
 ground state of the mean-field Hamiltonian becomes
 ill-defined. It is possible to propose an \emph{ad hoc} modification
 of the HFP approximation that takes into account the hard-core
 constraint. In the limit $U\to\infty$, one must have
 $b^\dagger_i b_i\le 1$. This constraint  can be satisfied by imposing 
 $|\lambda|^2\le 1$ on the order parameter and using as the variational
 energy  Eq.~(\ref{HFP-approx}) with $U=0$. Then, one obtains that for
 $zV<t^2/(2K)$, there is a first order transition, with $\lambda=0$
 and $\delta=0$
 for $\mu<-zt+zV-t^2/(2K)$ and $\lambda=1$ and 
$\delta=zt/K$ for $\mu>-zt+zV-t^2/(2K)$. The modified HFP
 approximation suggests that first-order magnetostriction transition
 caused by the condensation of triplet excitations can be observed
 also in $d>1$ provided the spin-phonon interaction is strong enough. 
The preceding discussion suggests that in order to describe correctly
 the first order magnetostriction transition associated with the Bose
 condensation of the triplets, we need to treat exactly the hardcore
 interaction of the triplets. In the quasi-one dimensional system of
 Eq.~(\ref{eq:ladder-strongcoupl}), this can be done in principle
 exactly thanks to the integrability of the XXZ spin
 chain\cite{bethe_xxx}.    
The mean field theory of the magnetostrictive
transition then proceeds\cite{mattis63_magnetostriction} by
writing $\delta_{n,m}=\delta \forall n,m$ and minimizing the free
energy with respect to $\delta$. The parameter $\delta$ measures the
relative change of the lattice parameter $\delta \propto\Delta a/a$. 
In the most general case, 
the free energy of the general XXZ chain in
the presence of magnetostriction have to be obtained from the
Thermodynamic Bethe
Ansatz
(TBA)\cite{takahashi_strings,shiroishi_nlie_xxx,kluemper_heisenberg_thermo,kluemper_thermo_xxx}. 
However, given the complexity of the TBA equations, it is better to
 consider the simplest case of $\Delta=0$ i. e. $J_\parallel^z=0$
 (XX chain) as in the original theory of the
 spin-Peierls transition\cite{pincus_spin-peierls,beni72_spinpeierls}.
In that particular case, by the Jordan-Wigner (JW)
transformation\cite{jordan_transformation}:
\begin{eqnarray}
  \label{eq:JW-transf}
  T_n^+&=&a_n^\dagger \cos \left(\pi \sum_{m<n} a^\dagger_m
  a_m\right), \\
  T_n^z&=&a^\dagger_n a_n -1/2, 
\end{eqnarray}
 the chain can be
mapped to a chain of free fermions and the calculation of its
thermodynamics is much simplified. We expect that this mapping is not
going to affect the qualitative properties of the chain under
field. In fact, when the magnetization per site of the XXZ chain is
near saturation, i.e. when the density of triplet excitations created
on the ladders is low, the effect of the short range interaction
measured by $\Delta$ becomes small. This is seen in particular in the
behavior of the Luttinger exponent $K$ which goes to $1$ in that low
density limit\cite{haldane_luttinger}. Thus, we expect that the
magnetostriction effects that happen near the transition between the
singlet state and the polarized state will be properly described by
the model~(\ref{eq:ladder-strongcoupl}) with $\Delta=0$.  
However, for a quantitative approach to the full magnetization curve, 
the use of TBA equations would become necessary. 
Even then, the insight gained
from the study of the XX chain will be useful to devise the correct
treatment of the TBA equations combined with the self-consistency
condition on $\delta$. Returning to the XX chain,
after  the JW transformation, and going to Fourier space,
the Hamiltonian of the chain reads:
\begin{eqnarray}
  \label{eq:hamiltonian}
  H= - J (1-\delta) \sum_k \cos k a^\dagger_k a_k +\frac K 2
  \delta^2-h
  (a^\dagger_k a_k -1/2)
\end{eqnarray}
where  we have written $J=J_\parallel$ and $h=h'$  to simplify
the notation. The operator $a_k$ annihilates a fermion of momentum
$k$. To study the thermodynamics, we consider that the
system is under fixed pressure, and work with the Gibbs free
energy\cite{mattis63_magnetostriction}. The variational Gibbs free
energy reads:
\begin{eqnarray}
  \label{eq:gibbs_energy}
  G(p,T,h;\delta)=\frac{K} 2 \delta^2 + p\delta + \frac h 2 -T
  \int_{-\pi}^{\pi} \frac{dk}{2\pi}
  \ln(1+e^{-((1-\delta)\epsilon(k) -h)/T}),
\end{eqnarray}
\noindent where $P$ is the pressure, 
$p=\eta P$ is the reduced pressure and we have units in which
$\hbar=k_B=1$. The
value of $\delta$ is obtained by minimizing $G(p,T,h;\delta)$ with
respect to $\delta$. This yields the equation:
\begin{eqnarray}
  \label{eq:delta-implicit}
  K\delta(p,T,h)+p=\int_{-\pi}^{\pi} \frac{dk}{2\pi}
  \frac{\epsilon(k)}{e^{[(1-\delta(p,T,h))\epsilon(k)-h]/T}+1},
\end{eqnarray}
\noindent and the Gibbs free energy is then ${\cal
  G}(p,T,h)=G(p,T,h;\delta(p,T,h))$. In the following sections, we
will study the Eq.~(\ref{eq:delta-implicit}) at zero temperature, and
exhibit the first-order magnetostrictive transition as a function of
the applied field. We will also
discuss the effect of the magnetostrictive effects on various
susceptibilities of the system. Then, we will turn to the effect of a
positive temperature.
 
\section{Zero temperature}
\label{sec:zeroT}
In this section, we first show that at $T=0$ and within mean field
theory, a first order phase transition as a function of the external
magnetic field is obtained in the array of 1D
XX chain coupled to acoustic phonons. Then, using a Landau-Ginzburg
expansion\cite{landau_statmech}, valid in the case of a stiff lattice,
we obtain analytically the expression of the discontinuity in the
magnetization and in the lattice parameter. This allows us to
represent the hysteresis cycle of the magnetization as a function of
the applied magnetic field. Finally, we discuss the anomalies in the
elastic constant and the magnetic susceptibility near the transition
as well as the magnetostriction coefficient. 

\subsection{Minimization of enthalpy at zero temperature}   
\subsubsection{First order transition}
\begin{figure}[htbp]
  \begin{center}
    \includegraphics[width=11cm]{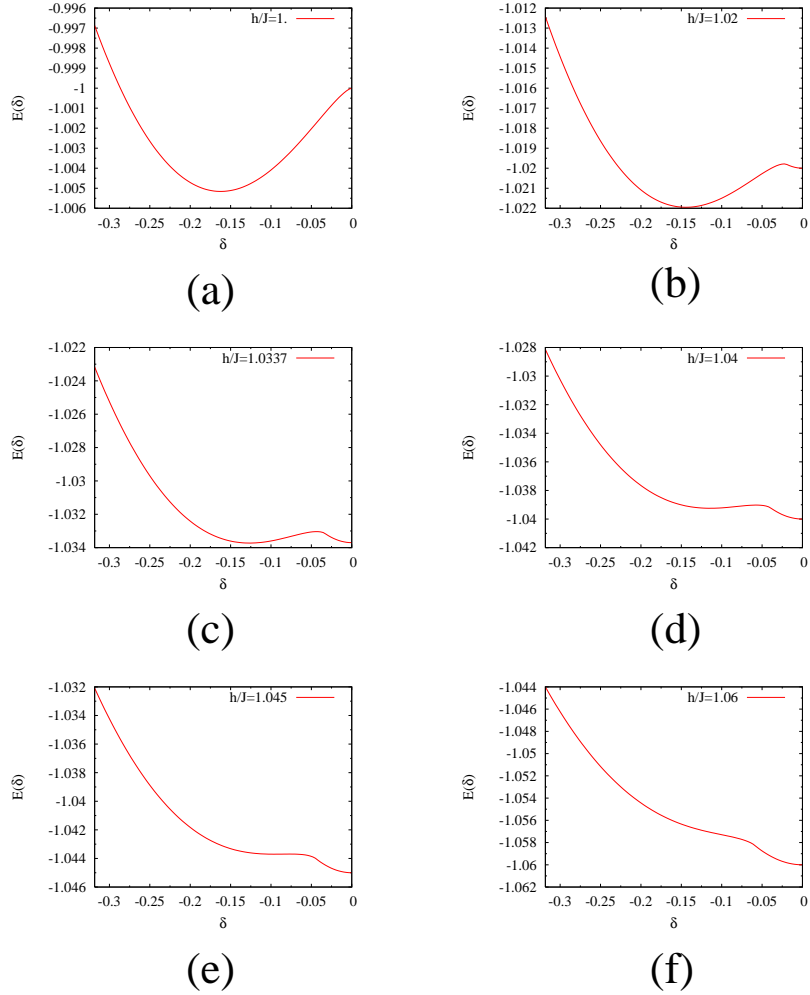}
    \caption{The ground state energy as a function of $\delta$ for
      different values of the magnetic field $h$. (a) For low field,
      there is a single minimum at $\delta<0$. (b) For a higher field
      $h>h_{c_<}$,
      a second minimum, at a higher value of the energy is
      present. This second minimum corresponds to a metastable state. 
      (c) For a critical value of the magnetic field, the two minimum
      become degenerate. (d) Above the critical field, the absolute
      minimum of the ground state energy is for $\delta=0$. A
      metastable minimum exists for $\delta<0$. (e) For $h=h_{c_>}$,
      the minimum at $\delta<0$ becomes unstable. (f) Above $h_{c_>}$,
      the only minimum is for $\delta=0$. In the range of
      $h_{c_<}<\delta<h_{c_>}$, hysteresis in $\delta$ as $h$ is
      varied is expected. }
    \label{fig:groundstate}
  \end{center}
\end{figure}

In order to study the equilibrium value of the relative change in the
lattice parameter $\delta$, we need an expression of the Gibbs
free energy as a function of $\delta$ valid for $T=0$.
At zero temperature, the Gibbs free energy (\ref{eq:gibbs_energy})
reduces to the enthalpy $H(p,h;\delta)=E_0(h;\delta)+p\delta$, where
$E_0$ is the ground state energy. The ground state of the Hamiltonian
(\ref{eq:hamiltonian}) is simply the Fermi sea of pseudofermions
filled up to the
chemical potential $h$. The Fermi wavevector of pseudofermions is
given by $-J(1-\delta)\cos k_F = h$, and the magnetization $M=\langle
T^z\rangle$ can be
expressed as a function of $k_F$ as:
\begin{eqnarray}
  \label{eq:magnet-kF}
  M=\frac {k_F}{\pi}-\frac 1 2.
\end{eqnarray}
Using the magnetization $M$ instead of the Fermi wavevector,
we can rewrite  the
ground state of the system as:
\begin{eqnarray}
  \label{eq:GS-energy}
  E(M,h,\delta)=-\frac{J}{\pi}(1-\delta) \cos (\pi M)
 + \frac K 2 \delta^2 -h M
\end{eqnarray}
The enthalpy must be minimized with respect to $\delta$ for fixed
$h,p$ giving:
\begin{eqnarray}
  \label{eq:system-a}
  K\delta + p +\frac J \pi \cos (\pi M)=0,
\end{eqnarray}
\noindent which is the zero temperature limit of
Eq.~(\ref{eq:delta-implicit}), and the magnetization is given by:
\begin{eqnarray}
\label{eq:system-b}
  J(1-\delta) \sin \pi M -h =0.
\end{eqnarray}
\noindent Both Eqs.~(\ref{eq:system-a})--~(\ref{eq:system-b}) are
valid only for $|M|<1/2$. For $M=\pm 1/2$, the kinetic energy of the
pseudofermions vanishes, and $E_0=K\delta^2/2-hM$. In this saturated
regime, one has obviously $\delta=-p/K$ independent of the
magnetization. In the regime of non-saturated magnetization, $\delta$
varies with the magnetization and the magnetic field, implying the
existence of magnetostrictive effects.
For $|M|<1/2$, it is possible to use  the equation
(\ref{eq:system-b}) to eliminate $M$ and write the ground state
energy as a function of $\delta$ only. We then obtain the final expression of
the ground state energy as:
\begin{eqnarray}
  \label{eq:energy-unsat}
   E_0(\delta,h)=-\frac{J}{\pi}\left[\sqrt{(1-\delta)^2-\left(\frac{h}{J}\right)^2}+\frac h J \arcsin\left(\frac{h}{J(1-\delta)}\right)\right] +K \frac{\delta^2}{2}
\end{eqnarray}
if $|h|<J|1-\delta|$ and: 
\begin{eqnarray}
  \label{eq:energy-sat}
  E_0(\delta,h)=\frac K 2 \delta^2 -h/2
\end{eqnarray}
if $|h|>J|1-\delta|$. The
Eq.~(\ref{eq:energy-unsat}) describes the XX chain not fully
polarized, whereas  Eq.~(\ref{eq:energy-sat})
describes  the fully polarized XX chain. the function $E_0(\delta,h)$
is plotted as a function of $\delta$ for different values of $h/J$ on
Fig.~\ref{fig:groundstate}. As is evident on
Fig.~\ref{fig:groundstate}, as $h$ is increased above $h_c$, 
the minimum existing
for $\delta=\delta_c<0$ acquires a higher energy than the minimum at
$\delta=0$ (see Figs.~\ref{fig:groundstate}(b)-(c)-(d)). As a result,
at $h=h_c$, the parameter $\delta$ jumps from $\delta_c$ to
$0$. Eq.~(\ref{eq:system-b}) then implies that the magnetization also
jumps from a value $M=M_c$ to $M=1/2$. In other words, the system
presents a first order transition at $T=0$ 
as a function of the magnetic field. In physical terms, 
the origin of this first order
transition is easy to grasp. When the system described by
(\ref{eq:ladder-array}) is in the spin gap state, it does not contain
any triplet. When the system starts to be polarized, the triplets first
occupy the states with the most negative kinetic energy. Due to the
hard-core repulsion, instead of being all in the lowest energy state,
they form a Fermi-sea.  By contracting, the lattice increases
the exchange energy and thus makes the kinetic energy of this Fermi
sea more negative. In
turn, this favors the condensation of a larger number of triplets in
the Fermi-sea, and
a further decrease  of the  energy of the triplets. 
This process is only limited
by the loss of elastic energy coming from the deformation of the
lattice. The natural consequence is that as soon as the magnetic field
overcomes the gap, a finite density of triplets appear. Due to the
fact that the triplets form a Fermi sea, their Bose condensation is
incomplete and this limitates the density of triplets. 
In the magnetized state, the system of coupled ladders possesses a
quasi-long range magnetic
order\cite{chitra_spinchains_field,furusaki_dynamical_ladder,giamarchi_coupled_ladders}
with: 
\begin{eqnarray}
  \label{eq:QLRO}
  \langle S_{n,m}^+ S_{n',m}^{-}\rangle &\sim&
  \frac{(-)^{n-n'}}{|n-n'|^{1/2}} +\frac{\cos [\pi(1-2\tilde{m})
  (n-n')]}{|n-n'|^{3/2}}, \\ 
 \langle S_{n,m}^z S_{n',m}^z\rangle &\sim& \frac{\tilde{m}^2}{4} + \frac{1}{|n-n'|^2} +
  \frac{\cos [2\pi \tilde{m}]}{|n-n'|^2}, 
\end{eqnarray}
\noindent $\tilde{m}=M+1/2$ being the magnetization per rung of the ladder. 
The power-law decay of the correlations of the staggered transverse
 magnetization~(\ref{eq:QLRO}) 
 can be interpreted a a quasi Bose
 condensation\cite{petrov04_bec_review} 
of the hardcore
bosons of the ladder in the lowest energy state of momentum
$(\pi,\pi)$. 

In fact, even in a quasi-one dimensional array of coupled chain, a
long range order of the BEC order parameter can develop. 
The reason is that in any real system, besides the spin-phonon
coupling, there exists
also an interladder exchange coupling $J'{\bf S}_{n,2m-1}\cdot {\bf
  S}_{2,2m+1}$. If this interladder coupling is
small enough compared to the rung coupling $J_\perp$, the singlet phase is stable\cite{gopalan_2ch}, and no antiferromagnetic
ordering is observed in the absence of an applied magnetic field. 
In the magnetized state however, this
interladder coupling gives rise at $T=0$ to a long range ordering of the
spins\cite{giamarchi_coupled_ladders}. In our problem, the Luttinger
exponent being equal to one, it is easily seen using the results of
Ref.~\onlinecite{giamarchi_coupled_ladders}  that the staggered
magnetization in the plane perpendicular to applied field behaves as
$(J'/J_\parallel)^{1/6}$. Therefore, one should also expect a
discontinuity of the staggered magnetization from $0$ to
$(J'/J)^{1/6}$ as the coupled ladder system becomes magnetized. This
discontinuity is the signature  of the BEC long range
order of the triplet bosons in the magnetized state.    

\subsubsection{Analytic study of the minimum of the enthalpy}
\begin{figure}[htbp]
  \begin{center}
    \includegraphics{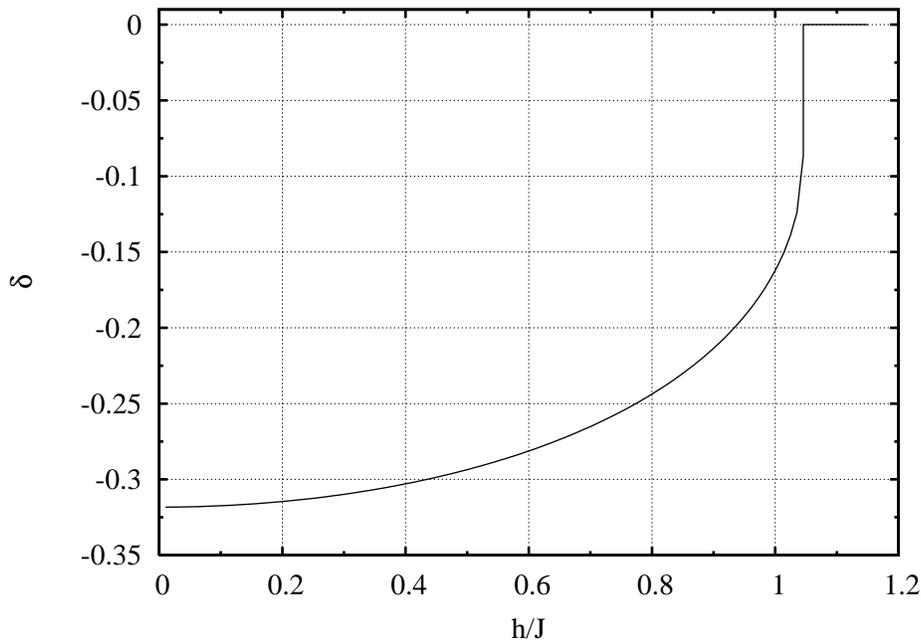}
    \caption{Striction for $T=0,P=0,K=1,J=1$. This is obtained from
      the solution of Eq.~(\ref{eq:4th-deg-delta}). It does not correspond to the
      thermodynamic striction which presents a jump for $h=h_c<h_{c_>}$.}
    \label{fig:striction-0K}
  \end{center}
\end{figure}

\begin{figure}[htbp]
  \begin{center}
    \includegraphics{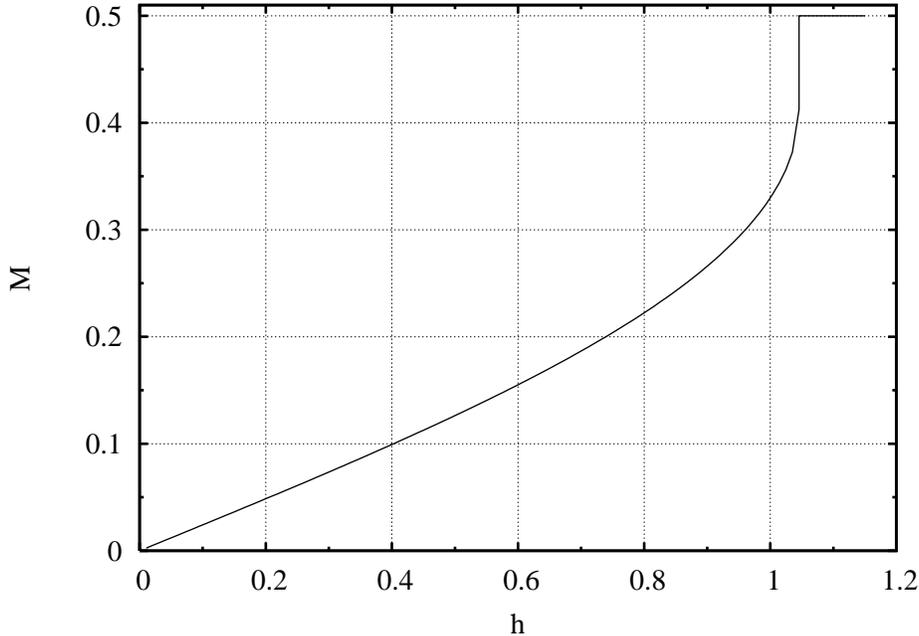}
    \caption{magnetization for $T=0,P=0,K=1,J=1$. This is obtained from
      the solution of Eq.~(\ref{eq:4th-deg-delta}) and this is not the
      thermodynamic magnetization.}
    \label{fig:magnetization-0K}
  \end{center}
\end{figure}

We now turn to an analytic study of the dependence of the parameter
$\delta$ on the magnetic field. 
In terms of $\delta$, the condition~(\ref{eq:system-a}) for
 the existence of the minimum
can be rewritten as:
\begin{eqnarray}
  \label{eq:min-delta}
   K\delta+p = -\frac 1 \pi \sqrt{J^2-\frac{h^2}{(1-\delta)^2}}
\end{eqnarray}
This equation is rewritten as the following equation of the fourth
degree:
\begin{eqnarray}
  \label{eq:4th-deg-delta}
    \delta^4 + 2\left(\frac p K -1\right) \delta^3 +\left[ 1-4\frac p K
    +\left(\frac p K\right)^2 -\left(\frac J{\pi K}\right)^2\right]
  \delta^2 + 2\left[\left(\frac J{\pi K}\right)^2 +\frac p K -
    \left(\frac p K\right)^2 \right] \delta +\frac  {h^2 -J^2} {(\pi
    K)^2} + \left(\frac p K\right)^2 =0.
\end{eqnarray}
It can be solved by the method of Cardan as exposed in
Ref.~\onlinecite{abramowitz_math_functions} p. 17, sec. 3.8.3. 
The dependence of $\delta$ on $h$
is represented on figure \ref{fig:striction-0K} and the resulting
magnetization on figure \ref{fig:magnetization-0K}. 

It can be shown that the lattice parameter given by the solution of
(\ref{eq:4th-deg-delta})  has a discontinuity as a
function of the magnetic field for a critical field $h_{c_>}$ given by:
\begin{eqnarray}
\label{eq:Hc2}
  \left(\frac{h_{c_>}}{J}\right)^2&=&\frac 7 8 + \frac 7 4 \frac p K + \frac 5
  8 \left(\frac p K\right)^2 +\left(\frac p K\right)^3 \nonumber \\
&+&\frac 1 4 \left(\frac{J}{\pi K}\right)^2 +
 \frac 1 4 \frac{\left(1+\frac p K\right)^3 +\left(\frac{J}{\pi K}\right)^2
   \left[8 +  \frac {6 p} K + 2 \left(\frac p K\right)^2\right]}
{1+\frac p K + \sqrt{\left(1+\frac p K\right)^2+8  \left(\frac{J}{\pi
        K}\right)^2}}.
\end{eqnarray}
For a field just below $h_{c_>}$, the value of $\delta$ is given by:
\begin{eqnarray}
  \label{eq:deltac2}
    \delta_{c_>} = -\frac p K +\frac 1 4 \left[1+\frac p K
    -\sqrt{\left(1+\frac p K\right)^2 + 8 \left(\frac{J}{\pi
          K}\right)^2}\right]
\end{eqnarray}
and above $h_{c_>}$, the value of $\delta$ is simply $\delta=0$ as
there are no physical solutions to the equation
(\ref{eq:4th-deg-delta}). The behavior of the enthalpy as a function
of $\delta$ for $h=h_{c_>}$ is represented on
Fig.~\ref{fig:groundstate}-(e). 

For $\delta=\delta_{c_>}$, the corresponding magnetization is
$M_{c_>}$ such that:
\begin{eqnarray}
  \label{eq:Mc2T0}
  \cos (\pi M_{c_>})=\frac{J}{\pi(K+p)+\pi \sqrt{(K+p)^2+8J^2}}. 
\end{eqnarray}
As with all first order transition,
 the solution of
Eq. (\ref{eq:4th-deg-delta}) may only describe a metastable state and
not the true minimum of the enthalpy, 
and the true thermodynamic transition happens
for a field $h_c$ lower than $h_{c_>}$. This is clearly seen on
Fig.~\ref{fig:groundstate}.  As a result,
the field $h_{c_>}$ only corresponds to an extremity of the hysteresis
cycle of the magnetization 
loop. The other extremity of the hysteresis cycle 
is obtained when the minimum at $\delta=0$ becomes unstable, i.e. when
the second derivative of
the enthalpy w.r.t. $\delta$ becomes negative for $\delta=0$. The
corresponding critical field is given by
$h_{c_<}=J(1+p/K)$. Obtaining analytically 
 the true thermodynamic critical field
is difficult in the general case. However, in the case of a stiff
lattice this can be done easily using a Landau-Ginzburg expansion.

\subsubsection{Landau-Ginzburg expansion at $T=0$}
 A more detailed description of the first order transition can be
obtained in the limit of very small magnetization jump, when it is
possible to expand the ground state energy in powers of $M-1/2$.
Expressing the ground state energy as a function of $M$ only,
using the Eqs.~(\ref{eq:GS-energy}) and (\ref{eq:system-a}), we
find:
\begin{eqnarray}
  \label{eq:GS-energy-M}
  E(M)=-\frac{J^2}{2\pi^2 K} \cos^2 (\pi M) - \frac J \pi
  \left(1+\frac p K\right) \cos (\pi M) -h M -\frac {p^2}{2K}
\end{eqnarray}
It is convenient to substract the energy of the fully polarized
system $E(M=1/2)=-h/2-p^2/(2K)$ to $E(M)$ and work with the
quantity:
\begin{eqnarray}
  \label{eq:E_0}
  E_0(M)=-\frac{J^2}{2\pi^2 K} \cos^2 (\pi M) - \frac J \pi
  \left(1+\frac p K\right) \cos (\pi M) -h \left( M -\frac 1 2\right)
\end{eqnarray}
Expanding Eq.~(\ref{eq:E_0}) to third order in $M-1/2$, we obtain:
\begin{eqnarray}
  \label{eq:LG-expansion}
  E_0(M)=-\frac{\pi^2 J}{6} \left( 1+\frac p K\right) \left(M-\frac 1
      2\right)^3 -\frac {J^2}{2K} \left(M-\frac 1 2\right)^2
    +\left[J\left(1+\frac p K\right)-h\right] \left(M-\frac 1
      2\right).
\end{eqnarray}
It is important to note that since $M<1/2$, the negative
coefficient of the cubic term does not lead to any instability.
Minimizing the ground state energy with respect to $M$, one finds:
\begin{eqnarray}
  \label{eq:min-condition-LG}
  \left[J\left(1+\frac p K\right)-h\right] -\frac {J^2}{K} \left(M-\frac 1
      2\right)+ \frac{\pi^2 J}{2} \left( 1+\frac p K\right) \left(M-\frac 1
      2\right)^2=0
\end{eqnarray}
For $h/J>(1+p/K)+J^2/[2\pi^2K(K+p)]\simeq h_{c_>}$, this equation has no
solution.  In this regime, $E_0(M)$ is a uniformly decreasing
function of $M$ and the minimum is obtained for $M=1/2$
corresponding to the fully saturated case. For $h/J<
(1+p/K)+J^2/[2\pi^2K(K+p)]$, the equation
(\ref{eq:min-condition-LG}) admits two solutions:
\begin{eqnarray}
  \label{eq:solutions-LG}
  M_{</>} &=&\frac 1 2 - \frac{\frac J K \pm \sqrt{\left(\frac J
        K\right)^2-2\pi^2\left[1+\frac p K -\frac h J\right](1+p/K)}}{\pi^2
    \left(1+\frac p K\right)}.
\end{eqnarray}
When $h/J<1+p/K=h_{c_<}$, the solution $M_>$  becomes larger than $1/2$
and thus unphysical . In this regime, $M_<$ is an absolute minimum
of $E_0(M)$. Let us first consider the intermediate regime
$h_{c_<}<h/J<h_{c_>}$. In this regime,
there is a minimum of $E_0(M)$ for $M=M_<$, a maximum for $M=M_>$
and a second minimum for $M=1/2$. The exchange of stability of
these two minima is at the origin of the first order transition.
We have $E_0(1/2)=0$, therefore, the stability of the minimum at
$M=M_<$ depends only on the sign of $E_0(M_<)$. Using the
equation~(\ref{eq:min-condition-LG}), we can rewrite:
\begin{eqnarray}
  E_0(M_<)=\left(M_<-\frac 1 2\right)^2
  \left[\frac{J}{2K}+\frac{\pi^3}{3} \left(1+\frac p K\right)
    \left(M_<-\frac 1 2\right)\right].
\end{eqnarray}
The stability of the two minima at $M=1/2$ and $M=M_<$ is thus
exchanged when:
\begin{eqnarray}
  \frac {J}{2K}=-\frac{\pi^2}{3}\left(1+\frac p
    K\right)\left(M_<-\frac 1 2\right).
\end{eqnarray}
At that point, one has:
\begin{eqnarray}
  \label{eq:mag-below-jump}
  M_<=\frac 1 2 -\frac{3J}{2\pi^2(K+p)}
\end{eqnarray}
and the magnetization jump at the transition is:
\begin{eqnarray}
  \label{eq:mag-jump}
  \Delta M= \frac{3J}{2\pi^2(K+p)}.
\end{eqnarray}
From this expression we immediately see that the condition of
applicability of the Landau-Ginzburg expansion is $J\ll (K+p)$,
that is equivalent to consider a very stiff lattice or a very high
pressure. Using the equation (\ref{eq:system-a}), it is also
possible to obtain an expression of the lattice parameter as:
\begin{eqnarray}
  \label{eq:lattice-LG}
  \delta=-\frac p K +\frac J K \left(M-\frac 1 2\right)=-\frac p K
  -\frac{3J^2}{2\pi K(K+p)} 
\end{eqnarray}
At the transition the jump of the lattice parameter is thus:
\begin{eqnarray}
  \label{eq:latt-jump}
  \Delta \delta = \frac J K \delta M = \frac{3 J^2}{2\pi^2
  K(K+p)}.
\end{eqnarray}

\begin{figure}[htbp]
  \begin{center}
    \includegraphics{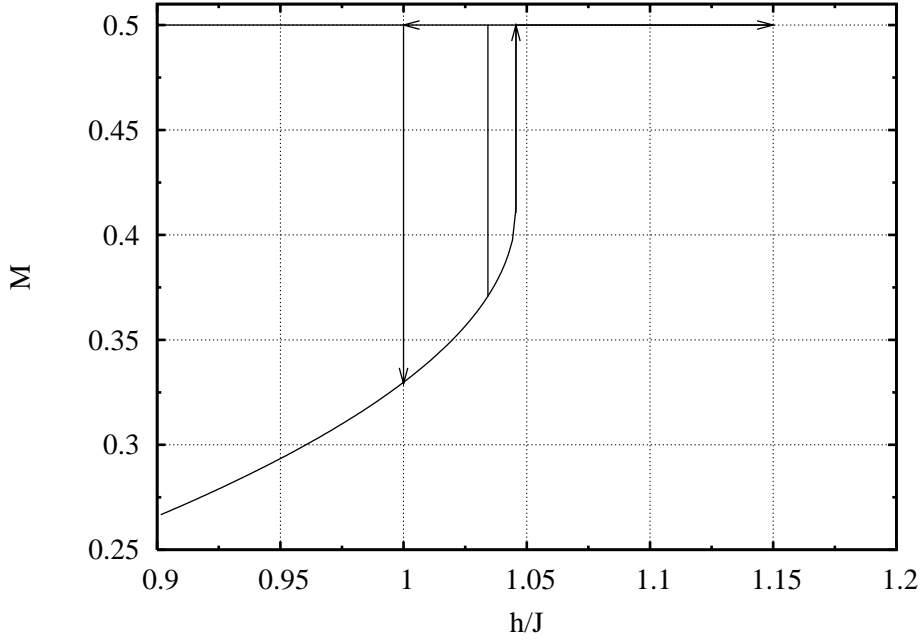}
    \caption{Hysteresis cycle for $T=0$ and $K=J=1,P=0$. The
      thermodynamic transition happens for $h=h_c\simeq 1.03J$, however
      the distorted state is metastable for
      $h_c<h<h_{c_>}\simeq1.045J$ and the undistorted 
state is metastable in the region
      $h_{c_<}=J<h<h_c$. As the field $h$ is increased from $0$ to
      $h_{c_>}$, the system remains on the metastable branch, and the
      magnetization jumps to $1/2$ at $h=h_{c_>}$. As the magnetic
      field is decreased, the magnetization drops from $1/2$ at
      $h=h_{c_<}$. This produces an hysteresis cycle in the
      magnetization. A similar hysteresis cycle is expected for the
      lattice distortion.}
    \label{fig:hysteresis}
  \end{center}
\end{figure}

The magnetic field at the transition is given by
Eq.~(\ref{eq:system-b}) as:
\begin{eqnarray}
  \label{eq:eq:hc-LG}
  \frac {h_c} J = 1+\frac p K + \frac {3 J^2}{8\pi^2 K(K+p)}, 
\end{eqnarray}
\noindent and obviously we have $h_{c_<}<h_c<h_{c_>}$. 
  As we have already
mentioned, in the intermediate regime, hysteresis can be observed
as there are two minimas, one of them being metastable. 
This is shown on  Fig.~\ref{fig:groundstate} where the energy for
$dE/dM=0$ becomes superior to the energy of the fully saturated
state at $h>h_c$. At that point the thermodynamic minimum becomes $M=1/2$,
and the branch with $dE/dM=0$ becomes metastable. When the field is
increased sufficiently rapidly, 
the magnetization may remain on the metastable branch until the field
reaches $h_{c_>}$. When the field is decreased from $h_{c_>}$, at
$h_c$ the solution with $M=1/2$ becomes metastable, and for a
sufficiently rapid decrease of the magnetic field, one can have a
magnetization $M=1/2$ until the magnetic field reaches $h_{c_<}$.  This is the
origin of the hysteresis cycle in the magnetization.  In
Fig.~\ref{fig:hysteresis} is shown a plot of the hysteresis cycle
of the magnetization for $T=0$. A similar hysteresis cycle exists for
the distortion and is  observed
in Fig. 3 of Ref.~\onlinecite{lorenzo04_cuhpcl_spinpeierls}, albeit in the
vicinity of $H=H_{c_1}$. In agreement with our picture, when the field
is decreased, the system remains on the metastable distorted state,
and when the distorted state becomes unstable the lattice parameter
jumps to its value in the undistorted state. When the field is
increased, the system remains in the metastable undistorted state
until it becomes unstable. 

Returning to our model, we note  that the
application of pressure is pushing the first order transition to
higher value of the applied magnetic field. This can be understood as a
consequence of the fact that in the distorted state $\delta<0$ and
$|\delta|$ is higher than in the undistorted state. As a result, the
stability of the distorted state is increased by the application of
pressure. Also, the hysteresis cycle becomes narrower as pressure is
increased. An analogous effect is observed in
ferromagnets\cite{grazhdankina69_1storder}, where application of
pressure reestablishes the second order Curie transition. It might be
interesting to investigate the effect of pressure on the transition
studied in Ref.~\onlinecite{lorenzo04_cuhpcl_spinpeierls} to see if the
behavior of $H_{c_1}$ as a function of pressure 
is in agreement with our prediction.

\subsection{Behavior of susceptibilities in the distorted phase}
\begin{figure}[htbp]
  \begin{center}
    \includegraphics{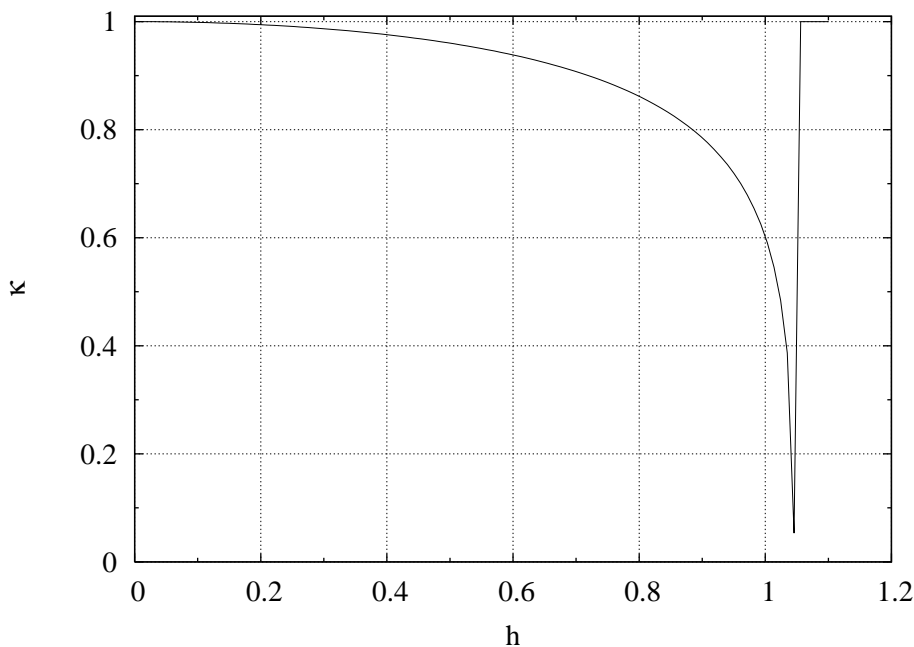}
    \caption{The effective elastic constant as a function of the
      magnetic field. Below the saturation field, $h_c$, the elastic
      constant is very sharply reduced. 
For $h=h_c$, the effective elastic constant   jumps to
      $K$. The curve is drawn for $J=K=1$. }
    \label{fig:elastic}
  \end{center}
\end{figure}

Now that we have described the first order transition, we would like
to discuss the effect of the spin-phonon coupling on the
susceptibilities of the system in the distorted phase. 
The following susceptibilities are of interest to us:\\
the compressibility:
\begin{eqnarray}
  \label{eq:compress-def}
  \tilde{\kappa}=-\left(\frac{\partial \delta}{\partial p}\right)_{h},
\end{eqnarray}
the magnetic susceptibility:
\begin{eqnarray}
  \label{eq:suscep-def}
  \chi=-\left(\frac{\partial M}{\partial h}\right)_{p},
\end{eqnarray}
\noindent and the parameter:
\begin{eqnarray}
  \label{eq:joule-del}
  \Lambda=-\left(\frac{\partial \delta}{\partial
      h}\right)_{p}=\left(\frac{\partial M}{\partial p}\right)_{h},
\end{eqnarray}
\noindent measuring the Joule
effect and its 
reciprocal, the Villari
effect\cite{bozorth51_ferromagnetism,bates39_magnetism}. The two
equivalent definitions of $\Lambda$ are direct consequences of the
definitions   $M=-\left(\frac{\partial
{\cal G}}{\partial h}\right)$, $\delta =\left( \frac{\partial {\cal G}}{\partial
p}\right)$.  
 The
equation of state results from the Eqs. (\ref{eq:system-a}-\ref{eq:system-b}).
Differentiating the second equation w.r.t. $p$, we find:
\begin{eqnarray}\label{eq:dmp-ddp}
  \left(\frac{\partial {M}}{\partial p}\right)_{h} = \frac{\tan(\pi M)}{\pi(1-\delta)}\left( \frac{\partial \delta}{\partial p}\right)_{h}
\end{eqnarray}
This equation indicates that applying a pressure for fixed
magnetic field induces a change of magnetization (Villari effect),
related to the
compressibility. Differentiating the first equation w.r.t. $p$,
instead gives:
\begin{eqnarray}
  K \left( \frac{\partial \delta}{\partial p}\right)_{h} + 1 -J
  \sin(\pi M) \left(\frac{\partial {M}}{\partial p}\right)_{h} = 0
\end{eqnarray}
Using Eq. (\ref{eq:dmp-ddp}), we finally obtain the following
expression for the compressibility:
\begin{eqnarray}
  \label{eq:compressibility-0K}
\frac 1 \kappa = -\left( \frac{\partial \delta}{\partial
p}\right)_{h} = \frac 1 {K-\frac{J}{\pi(1-\delta)}\tan(\pi M)\sin
(\pi M)}
\end{eqnarray}
We see that the compressibility of the system (which is in one
dimension the inverse of the elastic constant) is enhanced by the
interaction with the spin excitations. The physical reason is that
when the lattice spacing is diminished, the magnetic energy is
increased, which compensates for the loss of elastic energy. 
Expressed in terms of $\delta$, using Eq.~(\ref{eq:system-b}), 
the elastic constant reads:
\begin{eqnarray}
  \kappa=K-\frac 1{\pi} \frac{h^2}{(1-\delta)^2
    \sqrt{J^2(1-\delta)^2-h^2}}.
\end{eqnarray}
The behavior of $\kappa$ is represented on Fig.~\ref{fig:elastic}. 
It presents a minimum for $h=h_c$. 
The jump of the elastic constant from a low value to its original
value as the magnetic field is increased, implies a minimum in
elastic constant as a function of magnetic field. At zero
temperature such minimum appears more like a dip. Experimentally,
the presence of such a dip near the saturation field is known as
``$\Delta E$'' effect in ferromagnetic
materials\cite{bozorth51_ferromagnetism}. 
 Using Eqs.~(\ref{eq:eq:hc-LG}) and
(\ref{eq:lattice-LG}), we obtain that the minimum value of the elastic
constant reads for $J\ll K$: 
\begin{eqnarray}
  \label{eq:kappa-min}
  \kappa_c=K\left[\frac 1 3
    +\frac{29}{16}\left(\frac{J}{\pi(K+p)}\right)^2\right], 
\end{eqnarray}
\noindent showing that the elastic constant can be reduced to $1/3$ of
its value in the fully polarized phase. Applying pressure tends to
diminish further the value of $\kappa_c$. 
 The  behavior of $\kappa$ as a function of an applied magnetic field 
 has been recently
measured in a spin 1/2 Heisenberg chain material,  the coordination
polymer CuCCP\cite{wolf_sound_anomaly_spinchain}, via sound velocity
measurements. These experiments reveal  a pronounced minimum 
at the saturation field in qualitative
agreement  with the behavior represented on fig.~\ref{fig:elastic}. 
We will not attempt a quantitative comparison because the proper model
to use to study the anomaly in the elastic constant of CuCCP is the
Heisenberg model coupled to acoustic phonons and the thermodynamics of
this model, even in the mean field approximation requires a much more
sophisticated approach than the one of the present
paper\cite{kluemper_heisenberg_thermo,kluemper_thermo_xxx,shiroishi_nlie_xxx}.   

The most characteristic signature of magnetostriction is the
change  of magnetization under applied pressure. It is measured by:
\begin{eqnarray}
\Lambda=\left(\frac{\partial {M}}{\partial p}\right)_{h}= - \frac{\tan(\pi
     M)}
{\pi K (1-\delta) -J \tan(\pi M)\sin (\pi M)}=-\left(\frac{\partial
    {\delta}}{\partial h}\right)_{p}, 
\end{eqnarray}
\noindent  The behavior of
$\Lambda$ is reported on Fig.\ref{fig:alphaj}.
The application of pressure leads to a diminution of the
magnetization, and the application of a magnetic field in an increase
of the lattice parameter, which indicates that the system has a
positive magnetostriction\cite{bozorth51_ferromagnetism}. 
The variation of the magnetization $\Lambda$ is an
interesting quantity to measure, as well as $\left(\frac{\partial
{\delta}}{\partial h}\right)$, because they are non-zero only in
the phase that is not fully polarized, and their equality 
is the very signature of magnetostriction.

\begin{figure}[htbp]
  \begin{center}
    \includegraphics{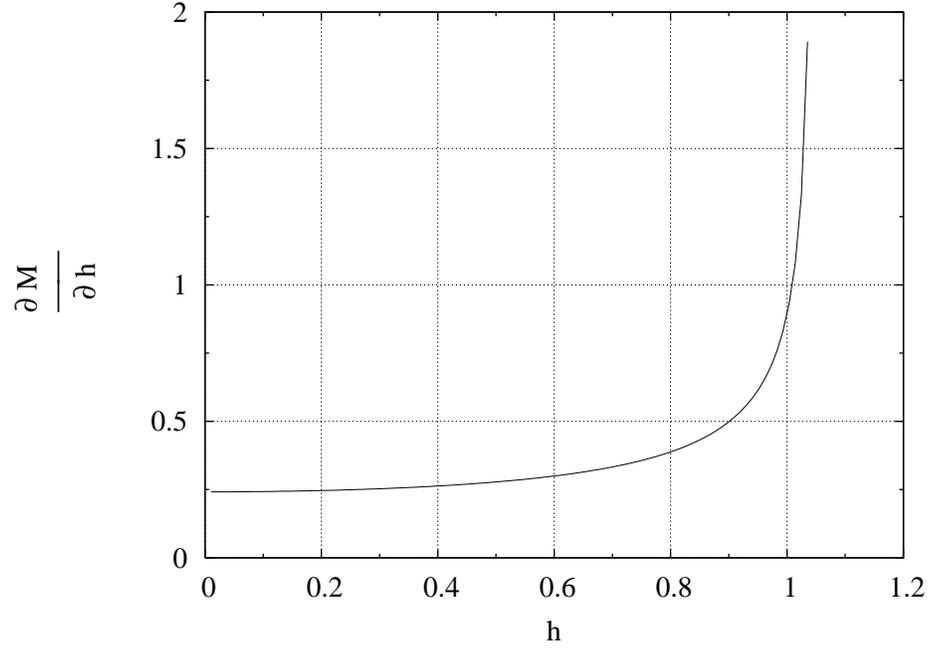}
     \caption{The magnetic susceptibility as a function of the
    magnetic field. It diverges at the critical field where the
    transition takes place.
    The plot is done for $J=K=1$. }
    \label{fig:susceptibility}
  \end{center}
\end{figure}

\begin{figure}[htbp]
  \begin{center}
    \includegraphics{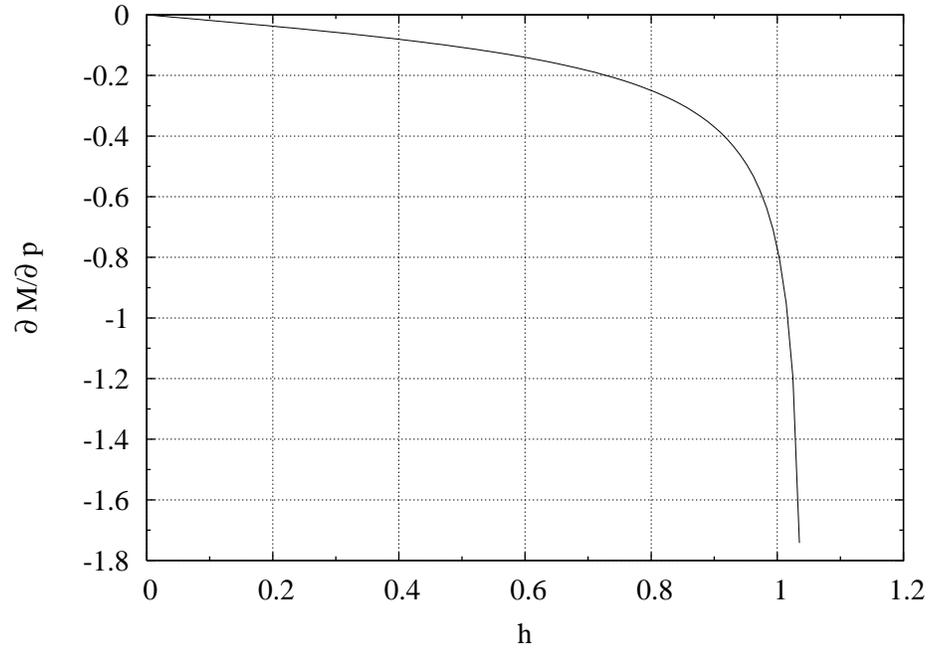}
    \caption{The Joule coefficient $\Lambda=\left(\frac{\partial M}{\partial
    p}\right)_h=-\left(\frac{\partial \delta}{\partial
    h}\right)_p$ as a function of the magnetic field.
    We see that applying pressure
     results in an decrease of the magnetization which indicates a
     positive magnetostriction. 
    The plot is done for $J=K=1$. }
    \label{fig:alphaj}
  \end{center}
\end{figure}

By considering the derivative of the equations of state
(\ref{eq:system-a}-\ref{eq:system-b}) w.r.t. $h$, we derive the
magnetic susceptibility. Differentiating the first equation gives:
\begin{eqnarray}
  \left(\frac{\partial \delta}{\partial h}\right)_p=\frac J K \sin
  (\pi M) \left(\frac {\partial M}{\partial h}\right)_p
\end{eqnarray}
Differentiating the second equation (\ref{eq:system-b}) then
yields for the susceptibility:
\begin{eqnarray}
\label{eq:susc}
  \chi = \left({\partial M}/{\partial h}\right)_p=\frac 1 {\pi J\left[
      (1-\delta) \cos (\pi M) -\frac J {\pi K} \sin^2(\pi
      M)\right]}.
\end{eqnarray}
The magnetic susceptibility can be rewritten as:
\begin{eqnarray}
  \chi = \chi_0(h,J(1-\delta)) \frac{K}{\kappa},
\end{eqnarray}
\noindent where $\chi_0(h,J)$ is the magnetic susceptibility of an XX
spin 1/2 chain with exchange constant $J$ in a magnetic field $h$. We
thus see that the susceptibility is strongly enhanced with respect to
the case of a perfectly rigid lattice. 
A plot of $\chi$ is reported in Fig.\ref{fig:susceptibility}. The
susceptibility has a maximum for $h=h_c$, and then falls to
zero. Using Eqs.~(\ref{eq:lattice-LG}) and (\ref{eq:eq:hc-LG}), we
find that the maximum susceptibility is:
\begin{eqnarray}
  \label{eq:chi-max}
  \chi_c=\frac{K}{J^2} \frac 3{1+\frac {189}{32}
    \left(\frac{J}{\pi(K+p)}\right)^2}, 
\end{eqnarray}
\noindent and thus the maximum value of the susceptibility is
proportional to the elastic modulus in the fully polarized phase. 
We also note that applying pressure results in an increase of the maximal
susceptibility. 
With the expressions of $\Lambda$ (Eq.~(\ref{eq:dmp-ddp}) and
$\chi$~(\ref{eq:susc}), we see that we have the
following relation:
\begin{eqnarray}
  (1-\delta)\frac \Lambda \chi=\frac{h}{K}.
\end{eqnarray}

\section{Positive temperature}
\label{sec:finiteT}

\begin{figure}[htbp]
  \begin{center}
    \includegraphics{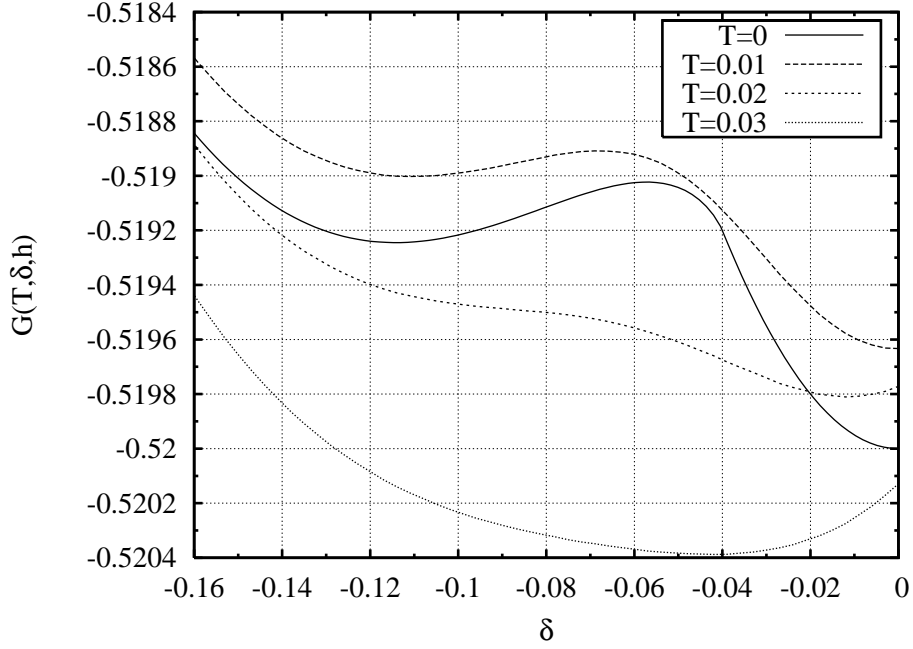}
    \caption{The evolution of the free energy as a function of
      temperature for $K=J=1$ and $h=1.04$. At zero temperature, the
      double minima  are neatly visible. The absolute minimum
      corresponds to $\delta=0$ indicating a non-magnetized phase,
      without deformation. For not too large positive temperature, the
      two minimum are still visible, but the barrier between them
      becomes smaller. Also, one notices that the minimum of the free
      energy corresponds to a solution with $\delta<0$, indicating a
      contraction of the lattice as $T$ increases, i.e. a negative
      dilatation coefficient.  As temperature is further increased,
      i.e.  for $T=0.03J$, there is only a single minimum indicating
      the disappearance of the first order transition.}
    \label{fig:energy-T}
  \end{center}
\end{figure}

At fixed magnetic field and fixed pressure, we can vary  the
temperature and the numerical analysis of the free energy as a
function of temperature shows that above a certain temperature, the
metastable minimum disappears, and one is left with a single minimum. 
This behavior
is illustrated for $P=0$ and $K=J$ on 
Fig.~\ref{fig:energy-T}.  
As a result,
above a certain temperature, the first order transition as a
function of the magnetic field disappears and is replaced by a
crossover. This is the analog of the critical point in the Van der
Waals mean field theory of the liquid-gas transition. In the present
problem, the role of the pressure is played by the magnetic
field. Applying the magnetic field is analogous to applying a pressure
to the Van der Waals fluid to liquefy it. It is important to point out
that the temperature at which the second minimum as a function of
$\delta$ disappears is a function of the applied magnetic field. The
second minimum disappears at lower temperature when its energy is
noticeably higher that the energy of the stable minimum. As a
result, the width of the hysteresis cycle decreases as a function of
temperature. A second effect of temperature is to blur the distinction
between the fully magnetized phase and the partially magnetized
phase. This effect can be seen on  Fig.~\ref{fig:energy-T}. As
temperature is increased, the stable minimum which corresponds to
$\delta=0$ is moved to $\delta<0$. Thus, as the magnetic field is
decreased for $T>0$, 
a reduction of the magnetization is observed even before the
magnetization jumps, in contrast with the behavior for $T=0$.

\subsection{Ginzburg-Landau expansion at finite T}

In this section, we turn to the derivation of the Ginzburg-Landau
expansion at finite temperature. Such expansion is derived from
the free energy:

\begin{eqnarray}
  F(T,\delta,h)=\frac K 2 \delta^2 -\frac 1 \beta \int \frac{dk}{2\pi} \ln
  (1+e^{-\beta[ (1-\delta)\epsilon(k)-h]}),
\end{eqnarray}
\noindent in the vicinity of $\delta=0$. Performing the Taylor
expansion in the logarithmic function, we find:

\begin{eqnarray}
  F(T,\delta,h)&=&-\delta  \int \frac{dk}{2\pi}
  \frac{\epsilon(k)}{e^{\beta(\epsilon(k)-h)}+1}+ \left( K-\beta \int \frac{dk}{8\pi}
  \frac{\epsilon(k)^2}{
    \cosh^2\left[\frac{\beta}{2}(\epsilon(k)-h)\right]} \right) \frac
{\delta^2}{2} \nonumber \\
 && - \beta^2 \int \frac{dk}{2\pi}
  \epsilon^3(k)
  \frac{\sinh\left[\frac{\beta}{2}(\epsilon(k)-h)\right]}{\cosh^3\left[\frac{\beta}{2}(\epsilon(k)-h)\right]} \frac{\delta^3}{24}
   +  \beta^2 \int \frac{dk}{2\pi}
  \epsilon^4(k) \frac{\cosh\left[\beta(\epsilon(k)-h)\right]
  -2}{\cosh^4\left[\frac{\beta}{2}(\epsilon(k)-h)\right]}\frac{\delta^4}{192}.
\end{eqnarray}
The solution with $\delta=0$ becomes unstable when:
\begin{eqnarray}
\label{eq:stab}
  K(T)=K - \beta \int_{-\pi}^{\pi} \frac{dk}{8\pi}
  \frac{\epsilon(k)^2}{\cosh^2\left[\frac{\beta}{2}(\epsilon(k)-h)\right]} <0.
\end{eqnarray}
At low temperature, we can estimate $K(T)$ as:
\begin{eqnarray}
  K(T)=K-\frac 1 \pi \left\{\frac{h^2}{\sqrt{J^2-h^2}} + \frac{\pi^2
      T^2}{6}  \frac{1}{\sqrt{J^2-h^2}} \left[ 2
      +\frac{h^2(J^2+2h^2)}{(J^2-h^2)^2}\right] \right\} 
\end{eqnarray}
Thus, the thermal correction can be neglected if:
\begin{eqnarray}
  \frac{\pi^2}{6} T^2 \left[ \frac 2 {h^2}+
    \frac{J^2+2h^2}{(J^2-h^2)^2}\right] \ll 1, 
\end{eqnarray}
This leads for $J\simeq h$ to:
\begin{eqnarray}
  T\ll \frac{8}{\pi^2} |J-h|
\end{eqnarray}
Except for a small region of size of order $\pi^2/8 T$ around $J$, 
the thermal correction to the stability of the undistorted solution is
negligible for low temperatures. This leads us to expect that the
first order transition persists for non-zero but sufficiently low
temperatures.  
We now show that there is a temperature $T_>$ above which the
first-order transition as a function of the magnetic field disappears.   
Using a straightforward minoration  $\cosh(x) \ge 1$ term in
Eq. (\ref{eq:stab}),
we find that:  
\begin{eqnarray}
\label{eq:inequality1}
  K-\frac{J^2}{8T}<K(T),
\end{eqnarray}
\noindent and when:
\begin{eqnarray}
  \label{eq:upboundTc}
  K>\frac{J^2}{8T},
\end{eqnarray}
\noindent i.e. $T>J^2/(8K)$, we have $K(T)>0$ and thus above a
temperature $T_{>}=J^2/(8K)$, there is certainly no first order
transition.  
As an aside, we note that 
the bound (\ref{eq:upboundTc}) can be improved when $h>J$. 
Then, $\cosh[(\epsilon(k)-h)/(2T)]<\cosh[(J-h)/(2T)]$  and
\begin{eqnarray}
\label{eq:inequality1bis}
  K-\frac{J^2}{8T\cosh^2\left(\frac{J-h}{2T}\right)}<K(T),
\end{eqnarray}
\noindent which shows that for low temperature, 
the solution with $\delta=0$ remains stable for high fields. We now
proceed to show that for $h<J$  
an upper bound for $K(T)$ can also be found in the regime of
$|h|<J$. Using this bound, we can show that the solution with
$\delta=0$ is unstable at low but finite temperature on a soft
lattice. This shows that the first order transition can be observed
for $T>0$ in that case. To obtain the bound, we use $k_F$ defined by
$\epsilon(k_F)=h$ and write:
\begin{eqnarray}
  \epsilon(k)-h = -J(\cos k-\cos k_F) = 2 J \sin
  \left(\frac{k-k_F}{2}\right) \sin \left(\frac{k+k_F}{2}\right)
\end{eqnarray}
We have for $k>0$ the obvious majoration:
\begin{eqnarray}
  |\epsilon(k)-h|<J|k-k_F|
\end{eqnarray}
Therefore:
\begin{eqnarray}
  \cosh^2 \frac{\epsilon(k)-h|}{2T} < \cosh^2 \frac{J|k-k_F|}{2T}, 
\end{eqnarray}
and:
\begin{eqnarray}
  K(T) < K-\int_0^\pi \frac{dk}{\pi T} \frac{J^2 \cos^2 k}{4 \cosh^2
    \frac{J|k-k_F|}{2T}} 
\end{eqnarray}
By a change of variables, this is rewritten as:
\begin{eqnarray}
\label{eq:inequality2}
  K(T)<K - \frac {2J} \pi \int_{-\frac{J}{2T}
    k_F}^{\frac{J}{2T}(\pi-k_F)} dx \frac{4 \cos^2(k_F+\frac{2T}{J}
    x)}{\cosh^2 x} 
\end{eqnarray}
In the limit $J k_F \gg T$ and $J(\pi-k_F) \gg T$, it is possible to
calculate approximately the upper bound as: 
\begin{eqnarray}
  K-\frac{J}{2 \pi} \left[ 1-\frac{\frac{2\pi T}{J}}{\sinh\frac{2\pi
        T}{J}}\cos 2k_F\right] +O(e^{-J/T\mathrm{Min}(k_F,\pi-k_F)}),   
\end{eqnarray}
\noindent which shows that when $K<J/(2\pi)$, the instability exists
for low temperature $T\ll J\mathrm{Min}(k_F,\pi-k_F)$. The first
order phase transition can therefore be observed at positive
temperature at least when the lattice is sufficiently soft. 

We now turn to an estimation of the temperature at which the first
order transition disappears valid in the case of a stiff lattice.
First, we express $K(T)$ using the density of states. 
We have:
\begin{eqnarray}\label{eq:K-T-dos}
  K(T)=K-\int_{-J}^J \frac{d\epsilon}{\pi \sqrt{J^2-\epsilon^2}}
  \frac{\epsilon^2}{4\cosh^2\left(\frac{\epsilon-h}{2T}\right)}
\end{eqnarray}
The divergence of the density of state for $\epsilon\to J$ is
responsible for the instability. Therefore, it is legitimate, when
$|J-h| \ll J$ to expand the integrand in Eq.~(\ref{eq:K-T-dos}) 
for $\epsilon \to J$. We find:
\begin{eqnarray}
\label{eq:K-T-dos2}
  K(T)=K-\int^J \frac{d\epsilon}{\pi \sqrt{2J(J-\epsilon)}}  \frac{J^2}{4\cosh^2\left(\frac{\epsilon-h}{2T}\right)}
\end{eqnarray} 
Performing the change of variable $\epsilon = J -(J-h) x$, we can
rewrite the integral in Eq. (\ref{eq:K-T-dos2}) as:
\begin{eqnarray}
  K(T)=K-\frac{J^{3/2}}{\pi\sqrt{2(J-h)}} {\cal
    F}\left(\frac{J-h}{T}\right), 
\end{eqnarray}
\noindent where: 
\begin{eqnarray}
{\cal F}(y) = \int_0^\infty \frac{dx}{\sqrt{x}} \frac{y}{4 \cosh^2
    \left[\frac{y}{2}(1-x)\right]} 
\end{eqnarray}
It is easy to see that ${\cal F}(y\to \infty)=1$ and ${\cal F}(y\to 0)
\sim y^{1/2}$. 
As a result, at low temperature, $T\ll (J-h)$, we recover the zero
temperature result, and at high temperature $J\gg T\gg (J-h)$, we find 
that:
\begin{eqnarray}
  K(T)=K-\frac{J^{3/2}}{2\pi T^{1/2}},  
\end{eqnarray}
and if $T>T_c$ where:  
\begin{eqnarray}
\label{eq:tcross}
\frac{T_c}{J}= \frac{J^2}{(2\pi K)^2}
\end{eqnarray}
the solution $\delta=0$ becomes stable, indicating that the first
order transition disappears. 
We note that given the exponential temperature dependence of the
spin-Peierls transition temperature\cite{pincus_spin-peierls}, the
first order transition appears at a much higher temperature than the
spin-Peierls transition in the case of a stiff lattice.  
\subsection{Numerics}

By the minimization of the Gibbs free energy
(\ref{eq:gibbs_energy}) w.r.t. $\delta$, we have obtained
the T dependence of the striction and the magnetization.
\begin{figure}[htbp]
  \begin{center}
    \includegraphics{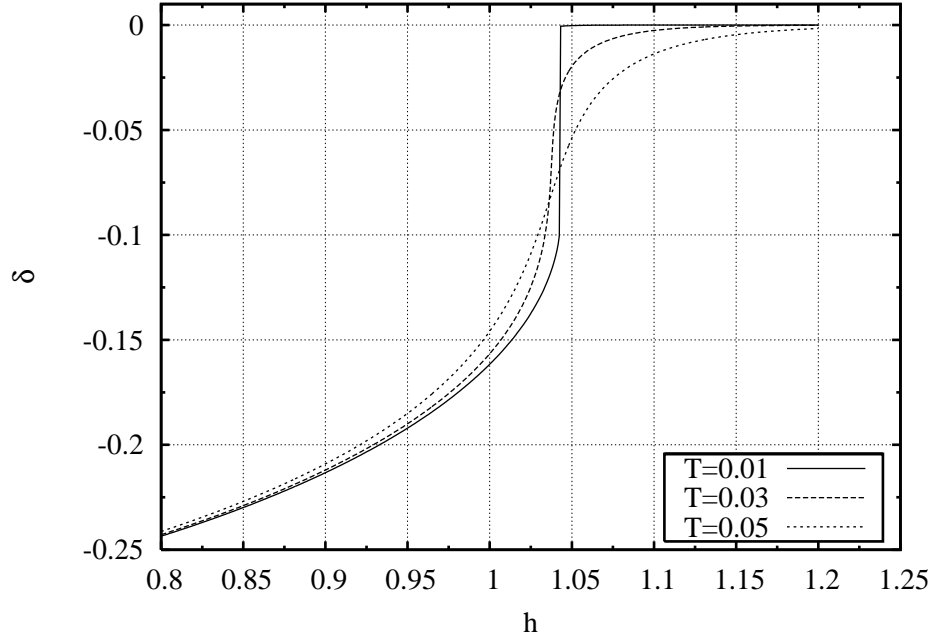}
    \caption{The striction as a function of the magnetic field for
      various temperatures and $K=J=1$. For $T=0.01$, the
      discontinuity becomes smaller than for $T=0$ but it
      is still present. For $T=0.03$ the discontinuity disappears, and
      is replaced by a rapid but continuous change of $\delta$. For
      $T=0.1$, the change becomes even more gradual. One can notice
      that $\delta$ reaches zero at higher field when the temperature
      is increased.}
    \label{fig:striction-T}
  \end{center}
\end{figure}

\begin{figure}[htbp]
  \begin{center}
    \includegraphics{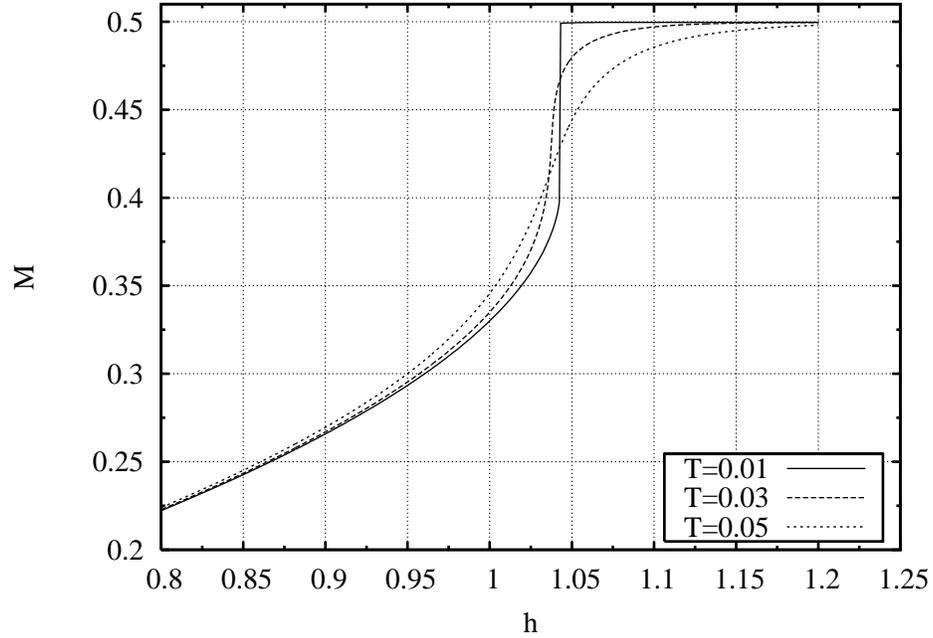}
    \caption{The magnetization as a function of the magnetic field for
      various temperatures and $K=J=1$. For $T=0.01$, the
      discontinuity is smaller, but it is still present. For $T=0.03$discontinuity disappears, and
      is replaced by a rapid but continuous change of $M$. As
      temperature is further increased, the change of $M$ becomes more
      gradual. One can notice that as $T$ is increased, the saturation
      field becomes higher. }
    \label{fig:magnetizion-T}
  \end{center}
\end{figure}

We find that the discontinuity in $\delta$ or $M$  becomes smaller
as $T$ is increased.  Also, we note there is a finite striction
and a  magnetization slightly below $1/2$ in the high $h$ phase.
The reason for this is that at finite temperature, pseudofermion
excitations are created which results in a lowering of the
absolute value of the kinetic energy-like term in
Eq.~(\ref{eq:delta-implicit}). As a result, the absolute value of
$\delta$ is diminished in the partially magnetized phase and
increased in the polarized phase. Above a certain temperature, the
discontinuity is replaced by a large slope, indicating a
continuous crossover from the fully polarized to the unpolarized
state. This behavior is reminiscent of the liquid-gas transition
in the Van der Waals fluid\cite{landau_statmech}. For $J=K=1$, we
find the transition temperature to be of order $0.03J$ in reasonable 
agreement with Eq.~(\ref{eq:tcross}).

\subsection{Susceptibilities at positive temperature}

For positive temperatures, the equation of state is obtained from the
equations:
\begin{eqnarray}
  \label{eq:state_T}
  K\delta+p=\int \frac{dk}{2\pi}
  \frac{\epsilon(k)}{1+e^{\beta[(1-\delta)\epsilon(k)-h]}} \\
  M=\int\frac{dk}{2\pi}\left [\frac 1 {1+e^{\beta[(1-\delta)\epsilon(k)-h]}}
  -\frac 1 2  \right]
\end{eqnarray}

\begin{figure}[htbp]
  \begin{center}
    \includegraphics{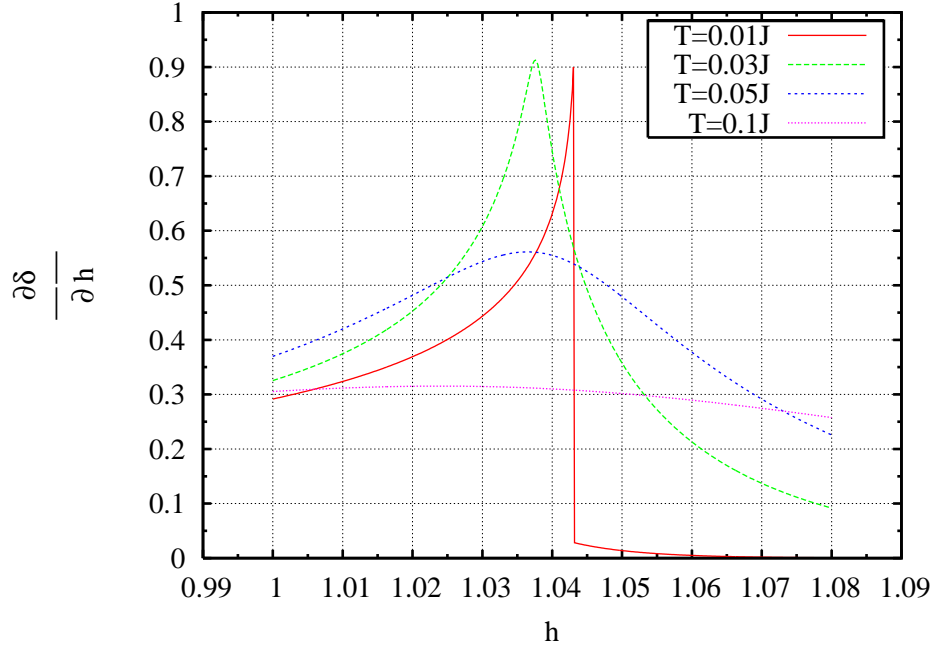}
    \caption{Derivative of the lattice parameter w.r.t. $h$ and
    varying the temperature.
    The plot is done for $J=K=1$. Below the transition temperature,
    the discontinuity is visible. Above the transition temperature,
    the discontinuity is replaced by a maximum. As the temperature
    increases, this maximum is lowered, and the region of anomaly is
    broadened.}
    \label{fig:ddeltah}
  \end{center}
\end{figure}

Taking the derivative of the first equation w.r.t. $h$ yields:
\begin{eqnarray}
\label{eq:ddelh}
  \left(\frac {\partial \delta}{\partial h}\right)_{T,p} = \frac{\beta \int \frac{dk}{2\pi}
  \frac{\epsilon(k)}{4 \cosh^2 \frac \beta 2 [(1-\delta)\epsilon(k)-h]
    }}{K- \beta \int \frac{dk}{2\pi}
  \frac{\epsilon(k)^2}{4 \cosh^2 \frac \beta 2 [(1-\delta)\epsilon(k)-h]
    }}= -\Lambda
\end{eqnarray}
The behavior of the Joule coefficient as a function of the magnetic
field at finite temperature is reported in Fig.\ref{fig:ddeltah}.
At very low temperatures it presents a cusp like behavior with a
jump at the critical field $h_c$. As the temperature increases the
cusp is replaced by a maximum for $h=h_M(T)<h_c$ 
that becomes broader at increasing
 $T$. 

\begin{figure}[htbp]
  \begin{center}
    \includegraphics{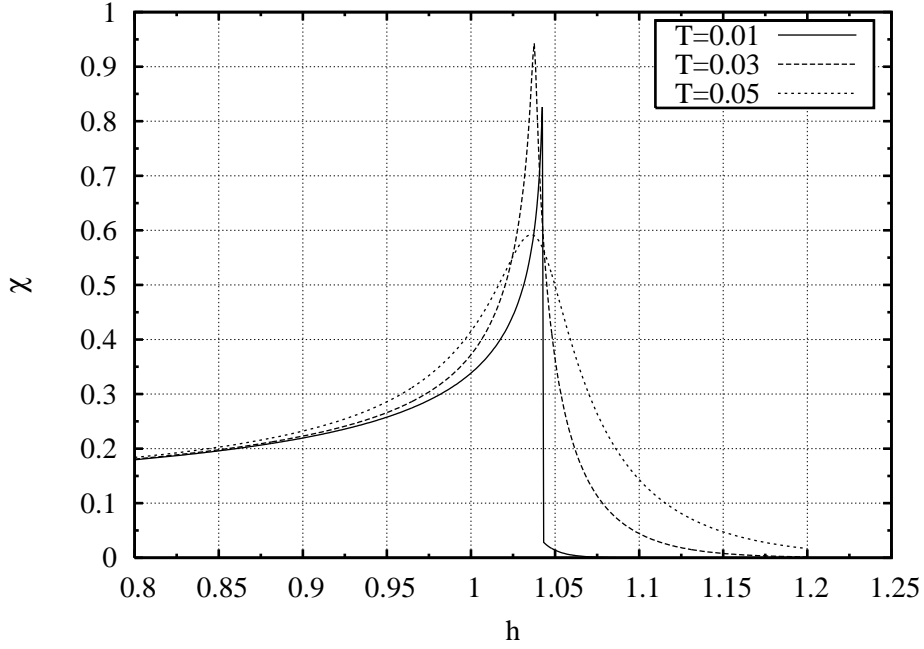}
    \caption{The magnetic susceptibility vs. $h$ for varying $T$. At
      low $T=0.01$, the magnetic susceptibility possesses a
      discontinuity. It is zero in the fully polarized phase, and
      becomes very large in the partially polarized state as the Fermi
      wavevector of magnetic excitations is nearing the band edge when
      magnetization is approaching saturation. For $T=0.03$, the
      discontinuity is replaced by a peak. As temperature is further
      increased, the height of the peak diminishes, while its breadth
      increases.}
    \label{fig:suscep-T}
  \end{center}
\end{figure}

Taking the derivative of the second equation w.r.t. $h$ then gives:
\begin{eqnarray}
 \left(\frac {\partial M}{\partial h}\right)_{T,p}= \beta \int \frac{dk}{2\pi}
  \frac{1}{4 \cosh^2 \frac \beta 2 [(1-\delta)\epsilon(k)-h]
    } + \frac{\left(\beta \int \frac{dk}{2\pi}
  \frac{\epsilon(k)}{4 \cosh^2 \frac \beta 2 [(1-\delta)\epsilon(k)-h]
    }\right)^2}{K- \beta \int \frac{dk}{2\pi}
  \frac{\epsilon(k)^2}{4 \cosh^2 \frac \beta 2 [(1-\delta)\epsilon(k)-h]
    }}
\end{eqnarray}
The plot of the magnetic susceptibility at finite temperature is
reported in Fig.~\ref{fig:suscep-T}. The singularity below $T_c$ is
changed into a maximum above $T_c$.

\begin{figure}[htbp]
  \begin{center}
    \includegraphics{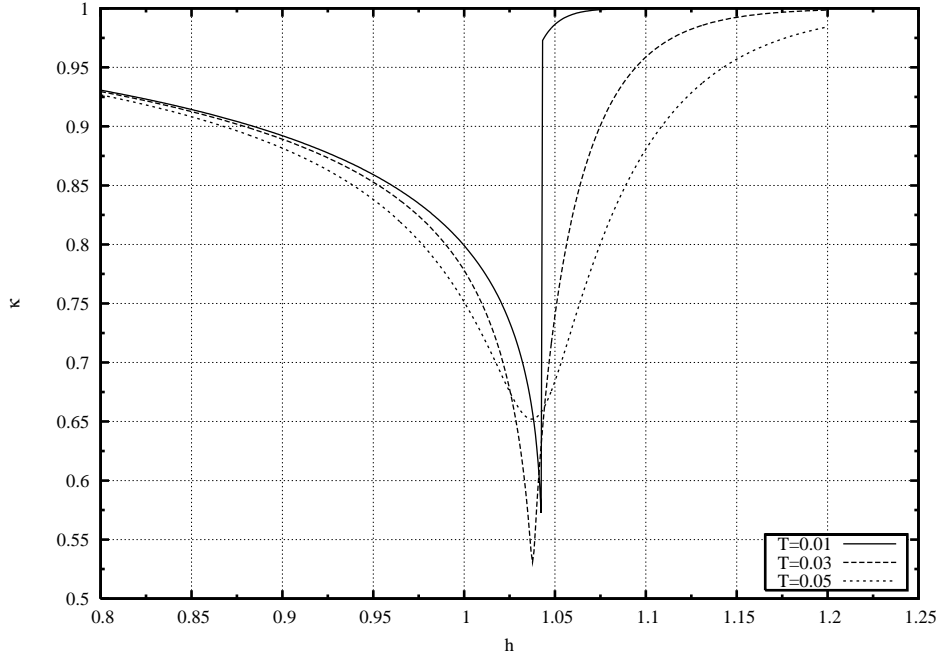}
    \caption{The effective elastic constant as a function of the
      magnetic field and different temperatures. The curve is drawn
      for $J=K=1$. We notice that in the fully polarized state at low
      temperature, the elastic constant quickly returns to its value
      in the absence of interaction with magnetic excitations. A jump
      of elastic constant is observed as the magnetic field is
      crossing the saturation value. At higher temperature, the jump
      is replaced by a minimum in the elastic constant. The minimal
      value of the elastic constant is an increasing function of
      $T$. We also note that the width of the region where the elastic
      constant decreases is larger as $T$ increases. }
    \label{fig:elasticT}
  \end{center}
\end{figure}

Similarly, taking the derivative of the first equation w.r.t. $p$
yields:
\begin{eqnarray}
 \frac {-1} \kappa=  \left(\frac {\partial \delta}{\partial p}\right)_{T,h} = \frac{-1}{K- \beta \int \frac{dk}{2\pi}
  \frac{\epsilon(k)^2}{4 \cosh^2 \frac \beta 2 [(1-\delta)\epsilon(k)-h]
    }},
\end{eqnarray}
At finite temperature the effective elastic constant is thus given by:

\begin{equation}
\label{eq:Keff} \kappa_T=K-\beta \int \frac{dk}{2\pi}
  \frac{\epsilon(k)^2}{4 \cosh^2 \frac \beta 2
  [(1-\delta)\epsilon(k)-h].
    }
\end{equation}
As shown in Fig.~\ref{fig:elasticT} the dip in the elastic
constant at $T=0$ (cf. Fig.~\ref{fig:elastic}) as a function if the
magnetic field  
is replaced by a minimum as the temperature $T>T_c$. 
The  the $T$ dependence of the elastic
constant is very reminiscent of the results reported in
Ref.~\onlinecite{wolf_sound_anomaly_spinchain} with the dip becoming broader
and less pronounced as the temperature is increased. 
However, in this reference,
the behavior of the elastic constant was only discussed in a
phenomenological way, by introducing an interaction proportional to
the striction and the square of the magnetization, i.e. a term
$-\delta (\langle S^z\rangle)^2$. A more quantitative approach could
be developed by treating the Hamiltonian
(\ref{eq:ladder-strongcoupl}) by a TBA mean-field theory.

Finally, taking the derivative of the second equation with respect to
$p$ allows one to check that Eq.~(\ref{eq:ddelh}) is recovered. 
\subsection{Other thermodynamic quantities: Dilatation coefficient and specific heat}

Besides the susceptibilities considered in the zero temperature
case, and recalculated at positive $T$, there are some other
quantities that can be defined for finite temperature, namely the
dilatation coefficient:
\begin{eqnarray}
  \label{eq:dilat-def}
  \alpha = -\left(\frac{\partial \delta}{\partial T}\right)_{p,h},
\end{eqnarray}
the specific heat:
\begin{eqnarray}
  \label{eq:specheat-def}
  C_p=-\left(\frac{\partial {\cal H}}{\partial T}\right)_{p,h}
\end{eqnarray}
\noindent where ${\cal H}$ is the enthalpy, and the quantity:

\subsubsection{Specific heat} \label{subs:specific_heat}

The internal energy is given by:

\begin{equation}
U(T,\delta,h)=\frac{\partial}{\partial \beta} \left (
\frac{F(T,\delta, h)}{T}\right )=\frac{K \delta^2}{2}+\int
\frac{dk}{2\pi}
\frac{[(1-\delta)\epsilon(k)-h]}{1+e^{\beta[(1-\delta)\epsilon(k)-h]}}.
\end{equation}

The specific heat is given by the derivative of the internal
energy, $C_V=\frac{\partial U}{\partial T}=-\frac{1}{T^2} \left(
\frac{\partial U}{\partial \beta}\right)$. Explicitly, the
specific heat at constant volume is given by:

\begin{equation}
\label{eq:cv} C_{V}(T,h)=-\frac{1}{ T^2} \int \frac{dk}{2\pi}
  \frac{[\epsilon(k)-h]^2}{4 \cosh^2 \frac \beta 2
  [\epsilon(k)-h].
    }
\end{equation}

\begin{figure}[htbp]
  \begin{center}
    \includegraphics{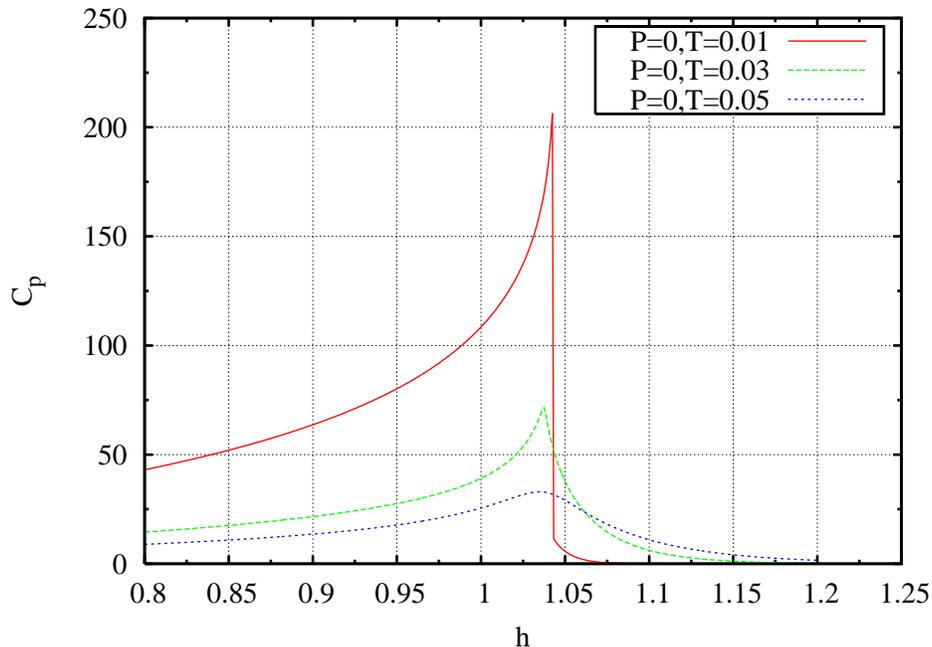}
    \caption{The specific heat for $T=0.01,0.03,0.05$ and $P=0$. We have
      $K=J=1$. The specific heat becomes very large near the
      transition. Above the transition temperature, a peak can be
      observed in the specific heat. The height of this peak
      diminishes as $T$ increases, whereas its breadth increases. }
    \label{fig:specheat}
  \end{center}
\end{figure}

To calculate the specific heat at constant pressure, we derive the
enthalpy ${\cal H}(t,\delta,h,p)=U(T,\delta,h)+p \delta$ and
evaluate $C_p(T,p)$ as:

\begin{equation}
\label{eq:cp} C_p=-\frac{1}{T^2} \int \frac{dk}{2\pi}
  \frac{[(1-\delta)\epsilon(k)-h]^2}{4 \cosh^2 \frac \beta 2
  [(1-\delta)\epsilon(k)-h]}
    +\frac 1 {T^3} \frac{\left(\int \frac{dk}{2\pi}
        \frac{\epsilon(k)\left[(1-\delta)\epsilon(k)-h\right]}{4
          \cosh^2\left( \frac {(1-\delta)\epsilon(k)-h}
            {2T}\right)}\right)^2}{K-\frac 1 T \int\frac{dk}{2\pi}
      \frac{\epsilon(k)^2}{4 \cosh^2\left( \frac {(1-\delta)\epsilon(k)-h}
            {2T}\right)}},
\end{equation}
\noindent where the second part comes from the variation of
$\delta$ with temperature. A plot of the specific heat is given on
Fig.~\ref{fig:specheat} for different temperatures. At very low-T
the specific heat presents a singularity that is replaced by a
maximum at finite temperature.

\subsubsection{Dilatation coefficient}

\begin{figure}[htbp]
  \begin{center}
    \includegraphics{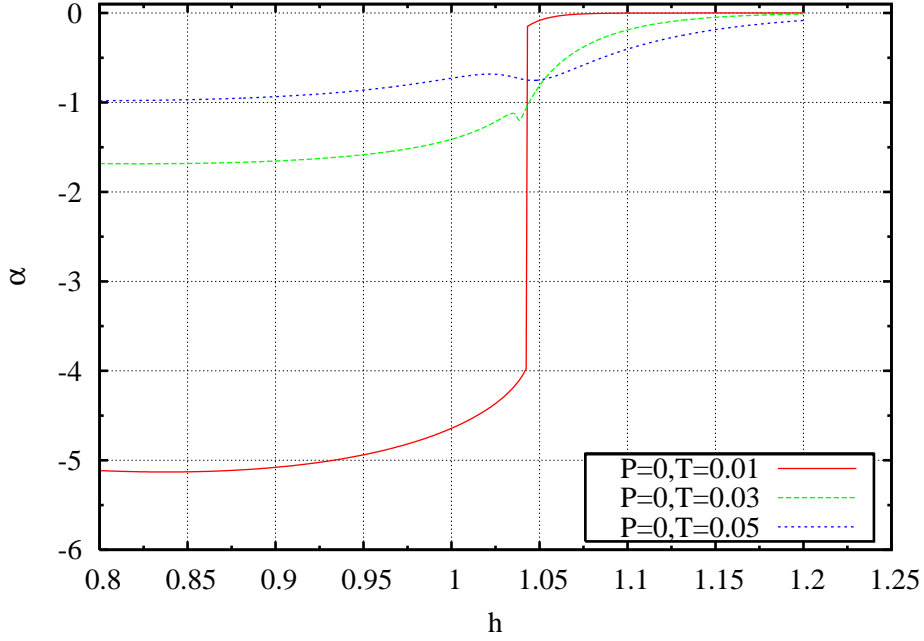}
    \caption{Dilatation coefficient for $T=0.01,0.03,0.05$ for $P=0$
      and $K=J=1$ as a function of the magnetic field. The dilatation
      coefficient vanishes in the fully polarized phase, and is
      negative in the partially polarized phase. We notice that the
      dilatation coefficient decreases strongly as temperature is
      increased. The behavior of the dilatation coefficient is
      non-monotonous as a function of the magnetic field.}
    \label{fig:dilat}
  \end{center}
\end{figure}

We use the definition of the dilatation
coefficient~(\ref{eq:dilat-def}), and Eq.~(\ref{eq:delta-implicit}) to
obtain:
\begin{eqnarray}
  \label{eq:dilat-result}
  \alpha=\frac 1 {T^2} \frac{\int \frac{dk}{2\pi} \frac
    {\epsilon(k)[(1-\delta)\epsilon(k)-h]}{4 \cosh^2\left(\frac
        {(1-\delta)\epsilon(k)-h}{2T}\right)}}{K-\int \frac{dk}{2\pi} \frac
    {\epsilon(k)^2}{4 \cosh^2\left(\frac
        {(1-\delta)\epsilon(k)-h}{2T}\right)}}
\end{eqnarray}
\noindent We note from Eq.~(\ref{eq:dilat-result}) 
 that the usual relation between $C_p$, $C_v$,  the
compressibility $\kappa$ and $\alpha$  (Eq. (16,9) of
Ref.~\onlinecite{landau_statmech}) is satisfied. 
The variation of the dilatation
coefficient with magnetic field is shown in Fig.\ref{fig:dilat}. As we
see, when the first order transition is present,  such
coefficient vanishes in the fully polarized phase and becomes
negative in the partially polarized phase. Above the temperature
$T_c$, the dilatation coefficient varies continuously, but its
behavior is non-monotonous when the magnetic field is close to the
value that induced the first-order transition at low temperature.

\section{Conclusions}
\label{sec:conclusions}

In this paper, we have discussed magnetostriction in an array of XX
spin chains coupled to three dimensional acoustic phonons. We have
shown that for low temperature, a first order transition would be
obtained as a function of the magnetic field, with a jump of both the
magnetization and the lattice constant. For higher temperatures, the
discontinuity is replaced by a crossover, but anomalies are still
visible in the elastic constant, the specific heat, the dilatation
coefficient  and the magnetic susceptibility. As we have explained in
the introduction, our results are also relevant to ladder systems
under magnetic field since the latter can be mapped in the limit of
strong rung coupling onto XXZ chain model under field. 
We stress again the important differences between magnetostriction and the
generalized spin Peierls transition of
Ref.~\onlinecite{nagaosa_lattice_ladder}. First, in the generalized spin
Peierls transition, a static lattice deformation of wavevector
$q=2\pi M$ where $M$ is the magnetization is
expected. This would lead to extra reflections in an X-Ray scattering
experiment. On the contrary, in magnetostriction, no superlattice is
formed, and no new reflections are expected. Second, in the
generalized spin Peierls transition, a spin gap is formed, leading to
an exponential suppression of the specific heat and magnetic
susceptibility at low temperature. In
the case of magnetostriction, no spin gap is formed, leading to a $T$
linear specific heat and a finite susceptibility 
at low temperature in the partially magnetized
phase. Also, in the case of the magnetostriction, anomalies are
expected in the elastic constant and the dilatation coefficient as a
function of the magnetic field. 
Of course, in a real system, both optical and acoustic phonons
are present, and one should expect a coexistence of magnetostriction
and generalized spin-Peierls
effects\cite{beni72_spinpeierls}. However, the first order transition
under applied field is observable at higher temperature than the
spin-Peierls transition so that the two effects can be separated in
principle.      
It would be interesting to analyze the more general case of
magnetostriction effects in a XXZ chain by using the
TBA\cite{takahashi_strings,kluemper_heisenberg_thermo,shiroishi_nlie_xxx,kluemper_thermo_xxx}
or by
using Quantum Monte Carlo method.   
We expect that the general
conclusions concerning the existence of the magnetostriction effect
and the possibility to separate it from the spin-Peierls transition
will not be affected, and a more quantitative agreement with
experiments will obtain. A more interesting extension of the present
work would be to consider a two dimensional system of hard bosons
coupled to phonons as this is relevant to the experiments on the
$\mathrm{TlCuCl_3}$ material\cite{sherman03_tlcucl3} and the CuHpCl
material \cite{stone02_cuhpcl}. 
In two dimensions, hard bosons on a lattice 
can be mapped onto fermions in a gauge
field\cite{fradkin_jw2d,azzouz93_jw2d}.  
For a low fermion density, the corresponding gauge field is small. 
Therefore, at low magnetization, it can be expected that one can
describe magnetostriction effect in a manner similar to the 1D case. 
We will leave this for future work.

\begin{acknowledgments}
E. O. thanks the Physics department of the University of Salerno
for hospitality in June 2003 and June 2004. E. O. and R. C. thank the
physics department of the University of Geneva for hospitality in
October 2003. We thank T. Giamarchi for discussions on the problem of
magnetostriction.  
\end{acknowledgments}

\appendix
\section*{Calculation of the field $h_{c_>}$} 

The Eq.~(\ref{eq:min-delta}) can be rewritten as:
\begin{eqnarray}
  \label{eq:polynome}
  P(\delta)=\left[\pi^2(K\delta+p)^2-J^2\right](1-\delta)^2 = h^2
\end{eqnarray}
The extrema of the function $P$ are given by $dP/(d\delta)=0$ i.e.:
\begin{eqnarray}
2(\delta-1)\left\{\delta^2 +(3p/K-1)\delta + \left[\left(\frac p K\right)^2
  -\frac p K -\left(\frac J {\pi K}\right)^2\right]\right\}=0,
\end{eqnarray}
giving a maximum for $\delta=\delta++$, with:
\begin{eqnarray}
  \delta_+=\frac 1 4\left[1-\frac{3p}{K}+\sqrt{\left(1+\frac p
        K\right)^2 + 8 \left(\frac J {\pi K}\right)^2}\right] 
\end{eqnarray}
and a minimum for $\delta=\delta_-$, with:
\begin{eqnarray}
  \delta_-=\frac 1 4\left[1-\frac{3p}{K}-\sqrt{\left(1+\frac p
        K\right)^2 + 8 \left(\frac J {\pi K}\right)^2}\right] 
\end{eqnarray}
The critical field $h_{c_>}$ is obtained by solving simultaneously 
the equation:
\begin{eqnarray}
\label{eq:zero-and-min}
  P(\delta_-)=h_{c_>}^2
\end{eqnarray}
This is done by introducing the polynomial:
\begin{eqnarray}
\tilde{P}(\delta) =  \delta^2 +(3p/K-1)\delta + \left[\left(\frac p K\right)^2
  -\frac p K -\left(\frac J {\pi K}\right)^2\right]
\end{eqnarray}
and writing
$P(\delta)-h_{c_>}^2=Q(\delta)\tilde{P}(\delta)+R(\delta)$. Since
$P(\delta_-)=h_{c_>}^2$ and $\tilde{P}(\delta_-)=0$, the equation
(\ref{eq:zero-and-min}) reduced to $R(\delta_-)=0$. Since $R$ is a
polynomial of degree 1, such an equation is trivial to solve. This
leads to Eq.~(\ref{eq:Hc2}). 


\begin{thebibliography}{10}

\bibitem{dalfovo99_bec_review}
F.~Dalfovo, S.~Giorgini, L.~P. Pitaevskii, and S.~Stringari,
\newblock Rev. Mod. Phys. {\bf 71}, 463 (1999),
\newblock and references therein.

\bibitem{affleck_field}
I.~Affleck,
\newblock Phys. Rev. B {\bf 41}, 6697 (1990).

\bibitem{tsvelik_field}
A.~M. Tsvelik,
\newblock Phys. Rev. B {\bf 42}, 10499 (1990).

\bibitem{chitra_spinchains_field}
R.~Chitra and T.~Giamarchi,
\newblock Phys. Rev. B {\bf 55}, 5816 (1997).

\bibitem{furusaki_dynamical_ladder}
A.~Furusaki and S.-C. Zhang,
\newblock Phys. Rev. B {\bf 60}, 1175 (1999).

\bibitem{usami_numerics_ladder}
M.~Usami and S.~Suga,
\newblock Phys. Rev. B {\bf 58}, 14401 (1998).

\bibitem{konik_spinfield}
R.~M. Konik and P.~Fendley,
\newblock Phys. Rev. B {\bf 66}, 144416 (2002).

\bibitem{lou00_magnetization_spin1}
J.~Lou, S.~Qin, T.-K. Ng, Z.~Su, and I.~Affleck,
\newblock Phys. Rev. B {\bf 62}, 3786 (2000).

\bibitem{hikihara_xxz}
T.~Hikihara and A.~Furusaki,
\newblock Phys. Rev. B {\bf 58}, R853 (1998).

\bibitem{haldane_bosons}
F.~D.~M. Haldane,
\newblock Phys. Rev. Lett. {\bf 47}, 1840 (1981).

\bibitem{cazalilla_1d_bec}
M.~A. Cazalilla,
\newblock J. Phys. B: At. Mol. Phys. {\bf 37}, S1 (2003).

\bibitem{petrov04_bec_review}
D.~Petrov, D.~Gangardt, and G.~Shlyapnikov,
\newblock J. Phys. IV {\bf 116}, 3 (2004).

\bibitem{giamarchi_coupled_ladders}
T.~Giamarchi and A.~M. Tsvelik,
\newblock Phys. Rev. B {\bf 59}, 11398 (1999),
\newblock cond-mat/9810219.

\bibitem{ho04_deconfined_bec}
A.~Ho, M.~Cazalilla, and T.~Giamarchi,
\newblock Phys. Rev. Lett. {\bf 92}, 130405 (2004).

\bibitem{katsumata_nenp_field}
K.~Katsumata et~al.,
\newblock Phys. Rev. Lett. {\bf 63}, 86 (1989).

\bibitem{chaboussant_cuhpcl}
G.~Chaboussant et~al.,
\newblock Phys. Rev. B {\bf 55}, 3046 (1997).

\bibitem{chaboussant_nmr_diaza}
G.~Chaboussant et~al.,
\newblock Phys. Rev. Lett. {\bf 80}, 2713 (1998).

\bibitem{chaboussant_mapping}
G.~Chaboussant et~al.,
\newblock Eur. Phys. J. B {\bf 6}, 167 (1998).

\bibitem{stone02_cuhpcl}
M.~B. Stone et~al.,
\newblock Phys. Rev. B {\bf 65}, 064423 (2002).

\bibitem{rice02_bec_magnons}
T.~M. Rice,
\newblock Science {\bf 298}, 760 (2002).

\bibitem{nikuni00_tlcucl3_bec}
T.~Nikuni, M.~Oshikawa, A.~Oosawa, and H.~Tanaka,
\newblock Phys. Rev. Lett. {\bf 84}, 5868 (2000).

\bibitem{oosawa02_kcucl3}
A.~Oosawa et~al.,
\newblock Phys. Rev. B {\bf 66}, 104405 (2002).

\bibitem{grenier04_cr3cr2br9}
B.~Grenier et~al.,
\newblock Phys. Rev. Lett. {\bf 92}, 177202 (2004).

\bibitem{jaime04_bacu2si2o6}
M.~Jaime et~al.,
\newblock Phys. Rev. Lett. {\bf 93}, 087203 (2004).

\bibitem{wessel01_spinliquid_bec}
S.~Wessel, M.~Olshanii, and S.~Haas,
\newblock Phys. Rev. Lett. {\bf 87}, 206407 (2001).

\bibitem{matsumoto02_tlcucl3}
M.~Matsumoto, B.~Normand, T.~M. Rice, and M.~Sigrist,
\newblock Phys. Rev. Lett. {\bf 89}, 077203 (2002).

\bibitem{matsumoto04_tlcucl3}
M.~Matsumoto, B.~Normand, T.~M. Rice, and M.~Sigrist,
\newblock Phys. Rev. B {\bf 69}, 054423 (2004).

\bibitem{misguich04_tlcucl3}
G.~Misguich and M.~Oshikawa,
\newblock Bose-einstein condensation of magnons in TlCuCl$_3$: Phase diagram and
  specific heat from a self-consistent Hartee-Fock calculation with a realistic
  dispersion relation,
\newblock cond-mat/0405422, 2004.

\bibitem{chubukov89}
A.~V. Chubukov,
\newblock JETP Lett. {\bf 49}, 129 (1989).

\bibitem{sachdev_bot}
S.~Sachdev and R.~N. Bhatt,
\newblock Phys. Rev. B {\bf 41}, 9323 (1990).

\bibitem{gopalan_2ch}
S.~Gopalan, T.~M. Rice, and M.~Sigrist,
\newblock Phys. Rev. B {\bf 49}, 8901 (1994).

\bibitem{popov_functional_book}
V.~N. Popov,
\newblock {\em Functional Integrals and collective excitations},
\newblock Cambridge University Press, Cambridge, 1987.

\bibitem{sherman03_tlcucl3}
E.~Y. Sherman, P.~Lemmens, B.~Busse, A.~Oosawa, and H.~Tanaka,
\newblock Phys. Rev. Lett. {\bf 91}, 057201 (2003).

\bibitem{vyaselev04_tlcucl3}
O.~Vyaselev, M.~Takigawa, A.~Vasiliev, A.~Oosawa, and H.~Tanaka,
\newblock Phys. Rev. Lett. {\bf 92}, 207202 (2004).

\bibitem{lorenzo04_cuhpcl_spinpeierls}
J.~E. Lorenzo et~al.,
\newblock Phys. Rev. B {\bf 69}, 220409(R) (2004).

\bibitem{calemczuk_heat_ladder}
R.~Calemczuk et~al.,
\newblock Eur. Phys. J. B {\bf 7}, 171 (1999).

\bibitem{nagaosa_lattice_ladder}
N.~Nagaosa and S.~Murakami,
\newblock J. Phys. Soc. Jpn. {\bf 67}, 1876 (1997).

\bibitem{pincus_spin-peierls}
P.~Pincus,
\newblock Solid State Commun. {\bf 4}, 1971 (1971).

\bibitem{cross_spinpeierls}
M.~C. Cross and D.~S. Fisher,
\newblock Phys. Rev. B {\bf 19}, 402 (1979).

\bibitem{bozorth51_ferromagnetism}
Bozorth,
\newblock {\em Ferromagnetism}, chapter~13,
\newblock Van Nostrand, Princeton, NJ, 1951.

\bibitem{vonsovskii74_magnetism2}
S.~V. Vonsovskii,
\newblock {\em Magnetism}, volume~II,
\newblock Wiley, New York, 1974.

\bibitem{bates39_magnetism}
L.~F. Bates,
\newblock {\em Modern Magnetism},
\newblock Cambridge University Press, Cambridge, UK, 1939.

\bibitem{grazhdankina69_1storder}
N.~P. Grazhdankina,
\newblock Usp. Fiz. Nauk {\bf 96}, 291 (1968),
\newblock [Sov. Phys. Usp. \textbf{11}, 727 (1969)].

\bibitem{wolf_sound_anomaly_spinchain}
B.~Wolf et~al.,
\newblock Phys. Rev. B {\bf 69}, 092403 (2004).

\bibitem{mila_ladder_strongcoupling}
F.~Mila,
\newblock Eur. Phys. J. B {\bf 6}, 201 (1998).

\bibitem{mattis63_magnetostriction}
D.~C. Mattis and T.~D. Schultz,
\newblock Phys. Rev. {\bf 129}, 175 (1963).

\bibitem{schultz_1dbose}
T.~D. Schultz,
\newblock J. Math. Phys. {\bf 4}, 666 (1963).

\bibitem{fisher_boson_loc}
M.~P.~A. Fisher, P.~B. Weichman, G.~Grinstein, and D.~S. Fisher,
\newblock Phys. Rev. B {\bf 40}, 546 (1989).

\bibitem{matsumoto04_magnetoelastic}
M.~Matsumoto and M.~Sigrist,
\newblock Ehrenfest relations and magnetoelastic effects in field-induced
  ordered phases,
\newblock cond-mat/0401411, 2004.

\bibitem{bethe_xxx}
H.~A. Bethe,
\newblock Z. Phys. {\bf 71}, 205 (1931).

\bibitem{takahashi_strings}
M.~Takahashi and M.~Suzuki,
\newblock Prog. Theor. Phys. {\bf 48}, 2187 (1972).

\bibitem{shiroishi_nlie_xxx}
M.~Shiroishi and M.~Takahashi,
\newblock Phys. Rev. Lett. {\bf 89}, 117201 (2002).

\bibitem{kluemper_heisenberg_thermo}
A.~{Kl\"umper},
\newblock Eur. Phys. J. B {\bf 5}, 677 (1998).

\bibitem{kluemper_thermo_xxx}
A.~{Kl\"umper} and D.~C. Johnston,
\newblock Phys. Rev. Lett. {\bf 84}, 4701 (2000).

\bibitem{beni72_spinpeierls}
G.~Beni and P.~Pincus,
\newblock J. Chem. Phys. {\bf 57}, 3531 (1972).

\bibitem{jordan_transformation}
P.~Jordan and E.~Wigner,
\newblock Z. Phys. {\bf 47}, 631 (1928).

\bibitem{haldane_luttinger}
F.~D.~M. Haldane,
\newblock Phys. Rev. Lett. {\bf 45}, 1358 (1980).

\bibitem{landau_statmech}
L.~D. Landau and E.~M. Lifshitz,
\newblock {\em Statistical Physics},
\newblock Pergamon, New York, 1960.

\bibitem{abramowitz_math_functions}
M.~Abramowitz and I.~Stegun,
\newblock {\em Handbook of mathematical functions},
\newblock Dover, New York, 1972.

\bibitem{fradkin_jw2d}
E.~Fradkin,
\newblock Phys. Rev. Lett. {\bf 63}, 322 (1989).

\bibitem{azzouz93_jw2d}
M.~Azzouz,
\newblock Phys. Rev. B {\bf 48}, 6136 (1993).

\end{thebibliography}

\end{document}